\documentclass{article}
\usepackage[top=0.75in, bottom=0.75in, left=0.75in, right=0.75in, margin=1in]{geometry} % Custom margins
\usepackage{amssymb}
\usepackage{amsmath}
\usepackage{graphicx}

\makeatletter
\expandafter\def\csname abx@macro@volume+number\endcsname{}
\makeatother
\usepackage[style=ieee]{biblatex}
\usepackage{mathtools}
\usepackage{booktabs}
\usepackage{siunitx}
\usepackage{subcaption}
\usepackage{multirow}
\usepackage{setspace}
\usepackage[percent]{overpic} % put this in the preamble

\usepackage{hyperref}
\usepackage{xcolor}
\definecolor{DarkRed}{RGB}{139,0,0}
\definecolor{DarkBlue}{RGB}{0,127,255}
\definecolor{Pink}{RGB}{255,51,51}
\usepackage{amsfonts}
\usepackage[outline]{contour}
\usepackage[scaled]{helvet}      % Helvetica
\usepackage{pdfrender}   % for thicker stroked text

\usepackage{algorithm}
\usepackage{algpseudocodex}
\usepackage{xcolor}

\usepackage{graphicx}
\usepackage{subcaption}

\usepackage{graphicx}
\usepackage{subcaption}

\usepackage{graphicx}
\usepackage{subcaption}
\usepackage[percent]{overpic}
\usepackage{xcolor}
\usepackage{tikz}
\usetikzlibrary{arrows.meta}

\graphicspath{%
    {figures/results_extra/}%
    {figures/time_profile/}%
    {figures/}%
}

\newcommand{\myarrow}[5]{%
    \put(#1,#2){%
        \makebox(0,0)[lb]{%
            \begin{tikzpicture}[x=\unitlength,y=\unitlength]
                \draw[-{Latex[length=6pt,width=7pt]}, red, line width=2pt]
                    (0,0) -- ({#3*#5},{#4*#5});
            \end{tikzpicture}%
        }%
    }%
}

\usepackage{etoolbox}

\newcommand{\ReconDashedLine}[1]{%
    \put(0,#1){%
        \tikz{%
            \draw[red!70,dashed,line width=1pt] (0,0) -- (5.27,0);%
        }%
    }%
}

\newcommand{\ReconTopCS}[4]{%
    \ifstrequal{#1}{us_0114}{%
        \begin{overpic}[#2,clip,width=\linewidth]{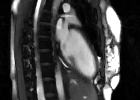}
            #3\ReconDashedLine{#4}
        \end{overpic}%
    }{}%
    \ifstrequal{#1}{us_9994}{%
        \begin{overpic}[#2,clip,width=\linewidth]{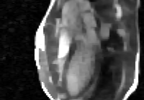}
            #3\ReconDashedLine{#4}
        \end{overpic}%
    }{}%
    \ifstrequal{#1}{us_0011}{%
        \begin{overpic}[#2,clip,width=\linewidth]{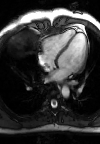}
            #3\ReconDashedLine{#4}
        \end{overpic}%
    }{}%
    \ifstrequal{#1}{us_9993}{%
        \begin{overpic}[#2,clip,width=\linewidth]{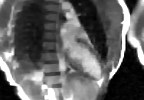}
            #3\ReconDashedLine{#4}
        \end{overpic}%
    }{}%
    \ifstrequal{#1}{us_8997}{%
        \begin{overpic}[#2,clip,width=\linewidth]{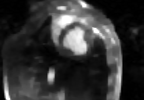}
            #3\ReconDashedLine{#4}
        \end{overpic}%
    }{}%
}

\newcommand{\ReconTopCineVN}[4]{%
    \ifstrequal{#1}{us_0114}{%
        \begin{overpic}[#2,clip,width=\linewidth]{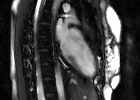}
            #3\ReconDashedLine{#4}
        \end{overpic}%
    }{}%
    \ifstrequal{#1}{us_9994}{%
        \begin{overpic}[#2,clip,width=\linewidth]{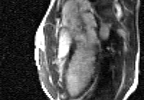}
            #3\ReconDashedLine{#4}
        \end{overpic}%
    }{}%
    \ifstrequal{#1}{us_0011}{%
        \begin{overpic}[#2,clip,width=\linewidth]{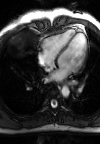}
            #3\ReconDashedLine{#4}
        \end{overpic}%
    }{}%
    \ifstrequal{#1}{us_9993}{%
        \begin{overpic}[#2,clip,width=\linewidth]{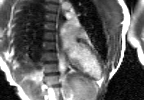}
            #3\ReconDashedLine{#4}
        \end{overpic}%
    }{}%
    \ifstrequal{#1}{us_8997}{%
        \begin{overpic}[#2,clip,width=\linewidth]{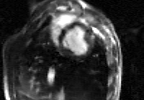}
            #3\ReconDashedLine{#4}
        \end{overpic}%
    }{}%
}

\newcommand{\ReconTopCineDiff}[4]{%
    \ifstrequal{#1}{us_0114}{%
        \begin{overpic}[#2,clip,width=\linewidth]{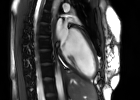}
            #3\ReconDashedLine{#4}
        \end{overpic}%
    }{}%
    \ifstrequal{#1}{us_9994}{%
        \begin{overpic}[#2,clip,width=\linewidth]{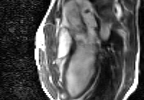}
            #3\ReconDashedLine{#4}
        \end{overpic}%
    }{}%
    \ifstrequal{#1}{us_0011}{%
        \begin{overpic}[#2,clip,width=\linewidth]{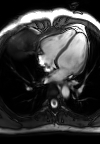}
            #3\ReconDashedLine{#4}
        \end{overpic}%
    }{}%
    \ifstrequal{#1}{us_9993}{%
        \begin{overpic}[#2,clip,width=\linewidth]{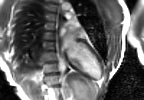}
            #3\ReconDashedLine{#4}
        \end{overpic}%
    }{}%
    \ifstrequal{#1}{us_8997}{%
        \begin{overpic}[#2,clip,width=\linewidth]{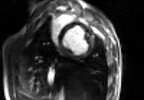}
            #3\ReconDashedLine{#4}
        \end{overpic}%
    }{}%
}

\newcommand{\TimeProfileCS}[2]{%
    \ifstrequal{#1}{us_0114}{%
        \begin{overpic}[width=\linewidth]{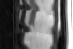}#2\end{overpic}%
    }{}%
    \ifstrequal{#1}{us_9994}{%
        \begin{overpic}[width=\linewidth]{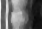}#2\end{overpic}%
    }{}%
    \ifstrequal{#1}{us_0011}{%
        \begin{overpic}[width=\linewidth]{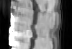}#2\end{overpic}%
    }{}%
    \ifstrequal{#1}{us_9993}{%
        \begin{overpic}[width=\linewidth]{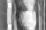}#2\end{overpic}%
    }{}%
    \ifstrequal{#1}{us_8997}{%
        \begin{overpic}[width=\linewidth]{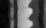}#2\end{overpic}%
    }{}%
}

\newcommand{\TimeProfileCineVN}[2]{%
    \ifstrequal{#1}{us_0114}{%
        \begin{overpic}[width=\linewidth]{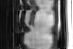}#2\end{overpic}%
    }{}%
    \ifstrequal{#1}{us_9994}{%
        \begin{overpic}[width=\linewidth]{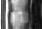}#2\end{overpic}%
    }{}%
    \ifstrequal{#1}{us_0011}{%
        \begin{overpic}[width=\linewidth]{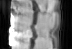}#2\end{overpic}%
    }{}%
    \ifstrequal{#1}{us_9993}{%
        \begin{overpic}[width=\linewidth]{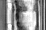}#2\end{overpic}%
    }{}%
    \ifstrequal{#1}{us_8997}{%
        \begin{overpic}[width=\linewidth]{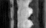}#2\end{overpic}%
    }{}%
}

\newcommand{\TimeProfileCineDiff}[2]{%
    \ifstrequal{#1}{us_0114}{%
        \begin{overpic}[width=\linewidth]{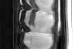}#2\end{overpic}%
    }{}%
    \ifstrequal{#1}{us_9994}{%
        \begin{overpic}[width=\linewidth]{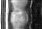}#2\end{overpic}%
    }{}%
    \ifstrequal{#1}{us_0011}{%
        \begin{overpic}[width=\linewidth]{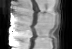}#2\end{overpic}%
    }{}%
    \ifstrequal{#1}{us_9993}{%
        \begin{overpic}[width=\linewidth]{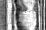}#2\end{overpic}%
    }{}%
    \ifstrequal{#1}{us_8997}{%
        \begin{overpic}[width=\linewidth]{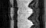}#2\end{overpic}%
    }{}%
}

\newcommand{\ReconTriplet}[9]{%
\begin{figure}[t]
    \centering

    \begin{subfigure}[b]{0.32\textwidth}
        \centering
        \subcaption*{\large\sffamily CS}
        \ReconTopCS{#1}{#5}{#6}{#9}
    \end{subfigure}
    \hfill
    \begin{subfigure}[b]{0.32\textwidth}
        \centering
        \subcaption*{\large\sffamily CineVN}
        \ReconTopCineVN{#1}{#5}{#6}{#9}
    \end{subfigure}
    \hfill
    \begin{subfigure}[b]{0.32\textwidth}
        \centering
        \subcaption*{\large\sffamily CineDiff}
        \ReconTopCineDiff{#1}{#5}{#6}{#9}
    \end{subfigure}

    \par\medskip

    \begin{subfigure}[b]{0.32\textwidth}
        \centering
        \TimeProfileCS{#1}{#8}
    \end{subfigure}
    \hfill
    \begin{subfigure}[b]{0.32\textwidth}
        \centering
        \TimeProfileCineVN{#1}{#8}
    \end{subfigure}
    \hfill
    \begin{subfigure}[b]{0.32\textwidth}
        \centering
        \TimeProfileCineDiff{#1}{#8}
    \end{subfigure}

    \par\medskip

    \includegraphics[width=\textwidth]{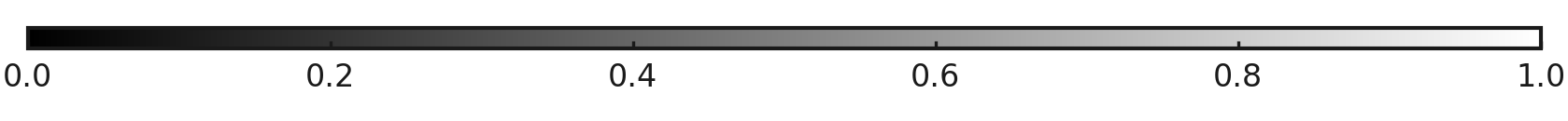}

    \caption{#3}
    \label{#4}
\end{figure}%
}

\usepackage{booktabs}
\usepackage[table]{xcolor}
\usepackage{multirow}

\newcommand\doubleplus{+\kern-1.3ex+}  % define our own \doubleplus command because stix and stix2 change the appearance of blackboard-bold symbols (\mathbb)

\renewcommand{\eqref}[1]{Equation~(\ref{eq:#1})}

\newcommand{\tabref}[1]{Table~\ref{tab:#1}} % MV: removed \!\!\!
\DeclareMathOperator*{\argmin}{arg\,min}

\renewcommand{\Hat}{\widehat}

\renewcommand{\vec}[1]{\ensuremath{\boldsymbol{#1}}}

\newcommand{\Hvec}[1]{\ensuremath{\Hat{\boldsymbol{#1}}}}

\newcommand{\Complex}{{\mathbb{C}}}

\newcommand{\footremember}[2]{%
    \footnote{#2}
    \newcounter{#1}
    \setcounter{#1}{\value{footnote}}%
}
\newcommand{\footrecall}[1]{%
    \footnotemark[\value{#1}]%
}
\let\svthefootnote\thefootnote
\newcommand\freefootnote[1]{%
  \let\thefootnote\relax%
  \footnotetext{#1}%
  \let\thefootnote\svthefootnote%
}

\contourlength{1.5pt}
\newcommand{\imagetextsize}{\fontsize{12pt}{12.5pt}}

\newcommand{\errlabel}{%
  \sffamily  \contour{black}{\textcolor{yellow}{\imagetextsize \hspace*{0em} {\bfseries x5} }}
}

\newcommand{\figuremetrics}[4]{%
  {\small\sffamily
  \begin{tabular}{@{}ll@{\hspace{0.8em}}ll@{}}
      ~~~~#1 & #2 \\
      ~~~~#3 & #4
  \end{tabular}%
  }%
}

\def\eqref#1{equation~\ref{#1}}
\def\1{\bm{1}}

\DeclareMathAlphabet{\mathsfit}{\encodingdefault}{\sfdefault}{m}{sl}
\SetMathAlphabet{\mathsfit}{bold}{\encodingdefault}{\sfdefault}{bx}{n}

\title{Patch-Based Diffusion Reconstruction for Accelerated Cardiac Cine MRI}
\author{%
    Xuan Lei\footremember{OSUece}{Department of Electrical and Computer Engineering, The Ohio State University, Columbus, OH, USA}%
    \and Philip Schniter\footrecall{OSUece}%
    \and Juliet Varghese\footremember{OSUbme}{Department of Biomedical Engineering, The Ohio State University, Columbus, OH, USA}%
    \and Rizwan Ahmad\footrecall{OSUbme}~\footrecall{OSUece}%
}
\date{}

\begin{document}

\maketitle
% \freefootnote{\mbox{}}
% \freefootnote{Corresponding author: Rizwan Ahmad (ahmad.46@osu.edu)}
\begin{NoHyper}
\freefootnote{\mbox{}}
\freefootnote{Corresponding author: Rizwan Ahmad (ahmad.46@osu.edu)}
\end{NoHyper}

\begin{center}
\vspace{5mm}
\end{center}

%%%%%%%%%%%
\begin{abstract}
\noindent\textbf{Purpose:} To develop and evaluate a diffusion-based reconstruction framework for highly accelerated 2D real-time (RT) cine cardiovascular magnetic resonance imaging (CMR).

\noindent\textbf{Methods:} We trained an unconditional patch-based diffusion model and incorporated it into a reconstruction framework, termed CineDiff, using diffusion posterior sampling for data consistency. CineDiff was evaluated in four settings: (i) 30 retrospectively undersampled breath-held cine at 1.5T and 3T from healthy participants across multiple acceleration rates, (ii) 15 prospectively undersampled free-breathing RT cine at 1.5T and 3T from patients indicated for clinical CMR, and (iii) 10 prospectively undersampled mid-field (0.55T) free-breathing scans, including five from healthy subjects and five from porcine models. For retrospective undersampling, reconstruction quality was assessed using peak signal-to-noise ratio (PSNR), structural similarity index measure (SSIM), learned perceptual image patch similarity (LPIPS), and deep image structure and texture similarity (DISTS). For prospective undersampling, image quality was evaluated by blinded expert scoring on a 5-point Likert scale.

\noindent\textbf{Results:} In retrospectively undersampled breath-held cine data, CineDiff achieved higher PSNR and SSIM and lower LPIPS and DISTS than the comparison methods across all evaluated acceleration rates. In prospectively undersampled free-breathing RT cine data, CineDiff received higher expert image-quality scores. Qualitatively, CineDiff reduced block-like artifacts and preserved finer anatomical detail compared with traditional compressed sensing and a variational network method, termed CineVN.

\noindent\textbf{Conclusion:} CineDiff enabled high-quality reconstruction of highly accelerated 2D RT cine CMR. The method also demonstrated robustness to out-of-distribution data, including mid-field and porcine acquisitions.

\end{abstract}

\subsection*{Abbreviations}
% 2D, two-dimensional; CS, compressed sensing; RT,  real-time; CineVN, variational network for cardiac cine; CMR, cardiovascular magnetic resonance imaging; E2E, end-to-end; GRO, golden ratio offset; ESPIRiT, eigenvalue approach to autocalibrating parallel MRI; PSNR, peak signal-to-noise ratio; R, acceleration rate; SSIM, structural similarity index measure; LPIPS, learned perceptual image patch similarity; DISTS, deep image structure and texture similarity.
% 2D, two-dimensional; CineVN, variational network for cardiac cine; CMR, cardiovascular magnetic resonance imaging; CS, compressed sensing; DISTS, deep image structure and texture similarity; E2E, end-to-end; ESPIRiT, eigenvalue approach to autocalibrating parallel MRI; GRO, golden ratio offset; LPIPS, learned perceptual image patch similarity; PSNR, peak signal-to-noise ratio; R, acceleration rate; RT, real-time; SSIM, structural similarity index measure.

2Ch, two-chamber; 2D, two-dimensional; 3Ch, three-chamber; 4Ch, four-chamber; bSSFP, balanced steady-state free precession; CineVN, variational network for cardiac cine; CMR, cardiovascular magnetic resonance imaging; CS, compressed sensing; DDIM, denoising diffusion implicit model; DDPM, denoising diffusion probabilistic model; DISTS, deep image structure and texture similarity; DPS, diffusion posterior sampling; E2E, end-to-end; ESPIRiT, eigenvalue approach to autocalibrating parallel MRI; GRO, golden ratio offset; LPIPS, learned perceptual image patch similarity; MRI, magnetic resonance imaging; PSNR, peak signal-to-noise ratio; R, acceleration rate; RT, real-time; SAX, short-axis; SNR, signal-to-noise ratio; SSIM, structural similarity index measure.

\section{Introduction}

%\textb{Cine CMR is acquired with breath-holding. Some patients can't hold their breath. Real-time (RT) cine enables free-breathing. RT cine relies on undersampling. RT cine reconstruction at high acceleration is challenging.}

Cardiovascular magnetic resonance imaging (CMR) is a well-established diagnostic tool. CMR-based cine is the gold standard for biventricular cardiac function assessment. In cine, the k-space data are typically acquired slice-by-slice using an electrocardiogram-triggered segmented sequence over multiple cardiac cycles, typically with breath-holds and parallel imaging methods~\cite{Pruessmann1999, Griswold2002}. 
However, breath-holding can be challenging for patients, especially those with advanced cardiopulmonary disease. Moreover, segmented acquisition fails in arrhythmic patients~\cite{nita2022real}, where the assumption of beat-to-beat consistency is violated.
Real-time (RT) cine CMR allows image acquisition during free-breathing and is robust to irregular heart rhythms, making it well-suited for such patient populations~\cite{hori2003rapid}.
To achieve sufficient spatial and temporal resolution for reliable assessment of cardiac function, RT cine relies on substantial k-space undersampling, which requires reconstruction methods that incorporate prior information about the images. Sparsity-based compressed sensing (CS)~\cite{Lustig2007}, which is now available on commercial scanners, has emerged as a popular choice for reconstructing RT images in clinical settings~\cite{sartoretti2019reduction}. 
While these conventional reconstruction methods can perform satisfactorily at low to moderate acceleration rates, recovering diagnostic-quality images at high acceleration remains challenging. At high acceleration rates, the conventional reconstruction methods often introduce spatial and temporal blurring~\cite{sharma2013clinical,contijoch2024future} and discrepancies in cardiac function assessment~\cite{craft2023comparison, yoon2019biases}. These limitations are expected to be more pronounced at mid-field scanners, which offer lower signal-to-noise ratio but are becoming more prevalent~\cite{kravchenko2025low} due to their cost-effectiveness.

Recently, a variety of deep learning–based reconstruction methods have been proposed for both static and dynamic MRI applications~\cite{knoll2020deep,heckel2024deep,safari2026advancing}. In particular, end-to-end (E2E) approaches, which directly take undersampled k-space (or an aliased image) as input and generate the final reconstructed image as output, have been extensively studied. Among E2E methods, unrolled networks have shown great promise~\cite{hammernik2018learning,aggarwal2018modl}. These networks reformulate iterative MRI reconstruction algorithms as deep networks, where each cascade mimics an optimization step that alternates between enforcing data consistency and applying learned image regularization. More recently, unrolled approaches have been extended to enable RT cine~\cite{CineVN}. However, like most E2E approaches, unrolled methods typically require training for specific sampling patterns and within a narrow range of acceleration factors to ensure acceptable reconstruction performance~\cite{gilton2021model}. In cine CMR, however, acceleration rates and other imaging parameters can vary substantially within the exam and across patients, which limits the practical flexibility of such methods.

Denoising diffusion models are powerful generative models that learn to sample from complex target distributions~\cite{ho2020denoising} and can be leveraged as data-driven priors to solve a wide range of inverse problems in imaging~\cite{kawar2022denoising}. Unlike traditional CS methods, diffusion-based MRI reconstruction does not rely on simple, handcrafted regularizers that do not fully capture the complex structure in the images. Unlike E2E approaches, diffusion-based reconstruction decouples the learned prior from the forward model, which allows a single trained model to be reused across different sampling patterns, acceleration rates, and acquisition settings. This flexibility is particularly attractive for cardiac MRI, where acquisition parameters often vary from scan to scan. Recent studies in brain and knee image reconstruction and enhancement have demonstrated that diffusion models improve image fidelity and robustness compared with conventional compressed sensing and supervised deep learning methods~\cite{chung2022score,ye2024decomposed,webber2024diffusion}.

% In this work, we propose a patch-based spatiotemporal diffusion framework for accelerated RT cine CMR, called CineDiff. Unlike prior diffusion-based MRI studies that operate primarily on full images or static frames, CineDiff models joint spatial-temporal structure using spatiotemporal patches and introduces a novel depatchification strategy that integrates patch-wise predictions within a measurement-guided reconstruction process. This design enables reconstruction that is free of patchification artifacts. While patch-based diffusion models have been explored in other modalities~\cite{hu2024learning}, their development in CMR remains limited. Preliminary results show that CineDiff outperforms existing CS and deep learning methods in image quality and, importantly, generalizes to free-breathing and low-field acquisitions despite being trained on high-SNR, breath-held segmented cine data.

In this work, we propose a patch-based spatiotemporal diffusion framework for accelerated RT cine CMR, called CineDiff. Unlike prior diffusion-based MRI studies that operate primarily on full images or static frames, CineDiff models joint spatial-temporal structure using spatiotemporal patches and introduces a novel de-patchification strategy that integrates patch-wise predictions within a measurement-guided reconstruction process. This design enables reconstruction that is free of patchification artifacts. While patch-based diffusion models have been explored in other modalities~\cite{hu2024learning}, their development in CMR remains limited. Preliminary results show that CineDiff outperforms existing CS and deep learning methods in image quality and, importantly, generalizes to free-breathing and mid-field acquisitions despite being trained on high-SNR, breath-held segmented cine data. We attribute this generalizability to the patch-based formulation of CineDiff, which reduces sensitivity to global spatiotemporal features.

\section{Methods}
\subsection{Cardiac MRI reconstruction}
In 2D dynamic MRI, including cine, the data are typically collected from multiple receive coils in the $k_x$-$k_y$-$f$ domain, where $k_x$-$k_y$ represent the spatial frequency grid, called k-space, and $f$ represents frames. To enable imaging with adequate temporal and spatial resolutions, each frame is often undersampled. In RT cine, k-space data from the $k^{\sf th}$ receive coil and the $f^{\sf th}$ frame can be represented by
\begin{equation}
    \vec{y}_{f,k}= \vec{P}_{f} \vec{\mathcal{F}} \vec{S}_{k} \vec{x}_f + \vec{\eta}_{f,k},
    \label{eq:forward_tk}
\end{equation}
where $\vec{y}_{f,k}\in \Complex^{M\times 1}$ is the measured data from the $k^{\sf th}$ receive coil and the $f^{\sf th}$ frame that has been vectorized into a column vector, $\vec{P}_f$ is an $M\times N$ binary mask for the $f^{\sf th}$ frame, $\vec{\mathcal{F}} \in \Complex^{N \times N}$ is the matrix representation of the discrete 2D Fourier transform, $\vec{S}_k \in \Complex^{N\times N}$ is the sensitivity map of the $k^{\sf th}$ coil, $\vec{x}_f\in \Complex^{N\times 1}$ is the image from the $f^{\sf th}$ frame that has been vectorized into a column vector, and $\vec{\eta}_{f,k} \in \Complex^{M\times 1}$ is zero-mean additive white Gaussian noise. For dynamic applications, the coil sensitivity maps are assumed to be time-invariant and pre-computed from calibration or time-averaged k-space data. The value of $N/M$ is often referred to as the acceleration rate, $R$.

By first vertically stacking $\vec{y}_{f,k}$, $\vec{S}_k$, and $\vec{\eta}_{f,k}$ across receive coils to build $\vec{y}_{f,1:K}\in \Complex^{KM\times 1}$, $\vec{S}_{1:K} \in \Complex^{KN\times N}$, and $\vec{\eta}_{f,1:K}\in\Complex^{KM\times 1}$, respectively, and then vertically stacking $\vec{y}_{f,1:K}$, $\vec{x}_f$, and $\vec{\eta}_{f,1:K}$ across frames as $\vec{y}_{1:F,1:K}\in\Complex^{KFM\times 1}$, $\vec{x}_{1:F}\in\Complex^{FN\times 1}$, and $\vec{\eta}_{1:F,1:K}\in\Complex^{KFM\times 1}$, respectively, and block-diagonally stacking $\vec{P}_f \vec{\mathcal{F}} \vec{S}_{1:K}$ across frames as $\vec{A}\in\Complex^{KFM\times FN}$, we arrive at $\vec{y}_{1:F,1:K}= \vec{A} \vec{x}_{1:F} + \vec{\eta}_{1:F,1:K}$. To improve readability, we drop subscripts $1:K$ and $1:F$, which leads to this forward model

\begin{equation}
    \vec{y}= \vec{A} \vec{x} + \vec{\eta}.
    \label{eq:forward}
\end{equation}

In RT cine, high values of $R$ imply that the problem in Equation \ref{eq:forward} is ill-posed and does not yield a satisfactory solution using least squares. The more common CS approach, which is available on most commercial scanners, solves Equation \ref{eq:forward} using regularized least squares, which leads to

\begin{equation}
    \Hvec{x}_{\sf cs}= \argmin_{\vec{x}} \|\vec{y}-\vec{A}\vec{x}\|_2^2 + \mathcal{R}(\vec{x}),
    \label{eq:cs}
\end{equation}
where $\mathcal{R}(\cdot)$ is a sparsity-promoting prior.

Recent deep learning methods have generally outperformed conventional CS in reconstruction quality~\cite{oscanoa2023deep}. Most deep learning approaches are supervised and train a network to map undersampled k-space data, or the corresponding aliased image, directly to the reconstructed image. More recently, unrolled methods have emerged as a particularly effective class of supervised deep learning reconstruction approaches. These methods mimic the iterative structure used to solve Equation~\ref{eq:cs} but replace handcrafted regularization updates, such as gradient-based or proximal steps, with learned network modules. %This formulation combines data consistency with learned image priors. 
The performance of such approaches, however, can degrade when the forward operator changes between the training and inference stages.

\begin{algorithm}[t]
\caption{CineDiff-based Reconstruction}
\label{alg:patch_based_CineDiff}
\begin{algorithmic}[1]
\Require $\vec{y}, \vec{A}, K = 1000, \text{DDIM~schedule~}k = 1\ldots K,
%\text{DDIM~schedule~} \{(\cdot)_k\}_{k=1}^{1000}, 
%\mathrm{DDIM~time~schedule}~\{I_k\}_{k=1}^{K},
\vec{x}_{\sf init}, T\leq 1000, \eta_{\mathrm{ddim}}$

\State $t \leftarrow T$
        \Comment{starting time step}

\If{$T<1000$} \Comment{warm start}
    \State $\vec{n} \sim \mathcal{N}(\vec{0}, \vec{I})$
    \State $\vec{p}_{t} \leftarrow
    \mathcal{P}\!\left(
    \vec{x}_{\sf init}
    + \sqrt{(1-\bar{\alpha}_{t})/\bar{\alpha}_{t}}\,\vec{n}
    \right)$
        \Comment{initial patches}
\Else 
    \State $\vec{p}_{t} \sim
    \mathcal{N}\!\left(
    \vec{0},
    \sqrt{(1-\bar{\alpha}_{t})/\bar{\alpha}_{t}}\,\vec{I}
    \right)$
        \Comment{initial patches}
\EndIf
\vspace{0.5em}
\For{$k = t, t-1, \dots, 1$}
    \State $\displaystyle
    \sigma_k \leftarrow
    \sqrt{(1-\bar{\alpha}_k)/\bar{\alpha}_k}$

    \State $\displaystyle
    \hat{\vec{p}}_{0|k} \leftarrow
    \mathcal{D}_{\vec{\theta}}(\vec{p}_k, \sigma_k)$
        \Comment{denoised patches}

    \State $\displaystyle
    \sigma_{k-1} \leftarrow
    \sqrt{(1-\bar{\alpha}_{k-1})/\bar{\alpha}_{k-1}}$

    \State $\displaystyle
    \vec{n}_k \sim \mathcal{N}(\vec{0}, \vec{I})$

    \State $\displaystyle
    \varsigma_k \leftarrow
    \eta_{\mathrm{ddim}}
    \sqrt{
    \frac{
    \sigma_{k-1}^2(\sigma_k^2-\sigma_{k-1}^2)
    }{
    \sigma_k^2
    }}$

    \State $\displaystyle
    h_k \leftarrow
    \sqrt{
    \frac{
    \sigma_{k-1}^2-\varsigma_k^2
    }{
    \sigma_k^2
    }}$

    \State $\displaystyle
    g_k \leftarrow 1-h_k$

    \State $\displaystyle
    \vec{p}'_{k-1} \leftarrow
    h_k\vec{p}_k
    + g_k\hat{\vec{p}}_{0|k}
    + \varsigma_k\vec{n}_k$
        \Comment{renoised patches}

    \State $\displaystyle
    \vec{p}_{k-1} \leftarrow \vec{p}'_{k-1}
    - \zeta_k \nabla_{\vec{p}_k}
    \left\|
    \vec{y} - \vec{A}\!\left(
    \smash[b]{
    \underbrace{
    \mathcal{P}^{-1}(\hat{\vec{p}}_{0|k})
    }_{\hat{\vec{x}}_{0|k}}
    }
    \right)
    \right\|_2^2$
        \Comment{data consistency update on patches}

    \Statex
\EndFor
\vspace{1.5em}
\State $\displaystyle
\hat{\vec{x}}_{0} \leftarrow \mathcal{P}^{-1}(\hat{\vec{p}}_{0|k})$
    \Comment{patches to image series}

\State \Return $\hat{\vec{x}}_{0}$

\end{algorithmic}
\end{algorithm}

\subsection{Diffusion models for MRI reconstruction}

% Denoising Diffusion Probabilistic Models (DDPM)~\cite{DDPM} aim to learn a data distribution $p_{\text{data}}(\vec{x}_0)$ in order to convert samples from a simple distribution (often Gaussian noise) into realistic images (or other data modalities, such as MRI images) that follows the data distribution. Denoising Diffusion Implicit Models (DDIM)~\cite{DDIM} are a deterministic (or partially stochastic) sampling variant of DDPM that enables faster inference by using fewer reverse-diffusion steps.

% Denoising diffusion probabilistic models (DDPM)~\cite{DDPM} learn a generative reverse process whose samples follow the data distribution $p_{\text{data}}(\vec{x}_0)$. 

Denoising diffusion models are powerful generative models that learn to sample from complex data distributions through a progressive denoising process, with denoising diffusion probabilistic models (DDPM)~\cite{ho2020denoising} serving as a canonical formulation. DDPM learns a reverse process that maps samples from a simple distribution, typically white Gaussian noise, to the target data distribution $p_{\text{data}}(\vec{x}_0)$, such as MR images. Denoising diffusion implicit models (DDIM)~\cite{DDIM} provides a deterministic or partially stochastic variant of this reverse process and enables faster sampling with fewer diffusion steps. Improving generation efficiency remains an active area of research, with recent advances in diffusion and related generative models enabling substantially faster generation~\cite{lu2022dpm,lipman2022flow,jiang2025fast}.

% DPS leverages the learned generative diffusion prior and incorporates the forward operator $\vec{A}$ and the measured data $\vec{y}$ to guide the denoising steps to solve inverse problems such as the accelerated cine MRI problem described in \eqref{eq:model-all-coils}. At each iteration, DPS refines the current result by guiding the results towards measurement $\vec{y}$ with a gradient step. As a result, DPS offers a way to couple powerful generative diffusion priors with measurement constraints, yielding high-quality reconstructions from undersampled data

Standard DDPM and DDIM sampling procedures do not by themselves enable consistency with the measured data. This has motivated a range of measurement-guided sampling methods that combine a learned diffusion prior with the forward model and acquired measurements to solve inverse problems. One such popular method is diffusion posterior sampling (DPS)~\cite{chung2022diffusion}, which, in our present setting for MRI, incorporates the forward operator $\vec{A}$ and measurements $\vec{y}$ to solve the accelerated cine reconstruction problem in Equation~\ref{eq:forward}. At each time step in the reverse diffusion process, DPS applies a diffusion-model update and then enforces measurement consistency through a gradient step derived from the data-fidelity term. This approach couples a strong learned image prior with the measured data through a known measurement model. Other guidance strategies have also been proposed for solving inverse problems~\cite{song2023pseudoinverse,zhu2023denoising,ye2024decomposed,zhang2025improving,bendel2025solving}.

% CineDiff uses a patch-based diffusion model as the prior and performs patch-based DPS with a DDIM-style sampler. The diffusion model is adapted from the video diffusion paper~\cite{video_diffusion_pytorch}. 
% The reconstruction process of CineDiff is illustrated in Fig.~\ref{fig:structure}, and the corresponding sampling procedure is summarized in algorithm~\ref{alg:patch_based_CineDiff}.
% For reconstruction, we initialize by sampling the required number of noise patches from a Gaussian distribution. Within the sampling loop, we first obtain a denoised estimate via the Tweedie estimator, then perform a DDIM update followed by a DPS-based data-consistency step. In the data-consistency step, the denoised estimate is depatchified and passed through the forward operator to compute the data-consistency gradient. After the final sampling iteration, we depatchify the reconstructed patches to obtain the final results. The patchifying and depatchifying steps are denoted as $\Pc(\cdot)$ and $\Pc^{-1}(\cdot)$, in Algorithm~\ref{alg:patch_based_CineDiff}. The operator $\Pc^{-1}(\cdot)$ uses ramp weighting in the overlapping regions to reduce edge artifacts introduced by patching. Details of these operators are provided in Appendix~\ref{app:patch_operations}.

\subsection{CineDiff framework}

In cardiac imaging, individuals with different body sizes and heart rates typically require different fields of view, spatial resolutions, and temporal resolutions.
Moreover, the content of the cine images is not oriented or centered consistently because of the wide variations in the body size and the relative location and orientation of the hearts. Therefore, compared to imaging of other organs, such as the brain, cardiac imaging presents more spatiotemporal variation in image content. 
In the face of this complex and rich data distribution and limited training data, we opted to train the model to recover spatiotemporal patches rather than full images. This approach alleviates data scarcity, because not only do patches represent a simpler distribution compared to the entire image but also a larger number of unique patches can be extracted from a single cine series. Another motivation for the use of patches is the inevitable training-testing mismatch, where training data are collected under breath-held conditions while testing data are collected under free-breathing conditions. We conjecture that, being a patch-based method, CineDiff is less sensitive to large-range (across several pixels) variations in the images. In contrast, patches capture the local spatiotemporal structure at the potential cost of ignoring long-range dependencies. 

\begin{figure}[h]
  \centering
  \includegraphics[width=0.91\linewidth]{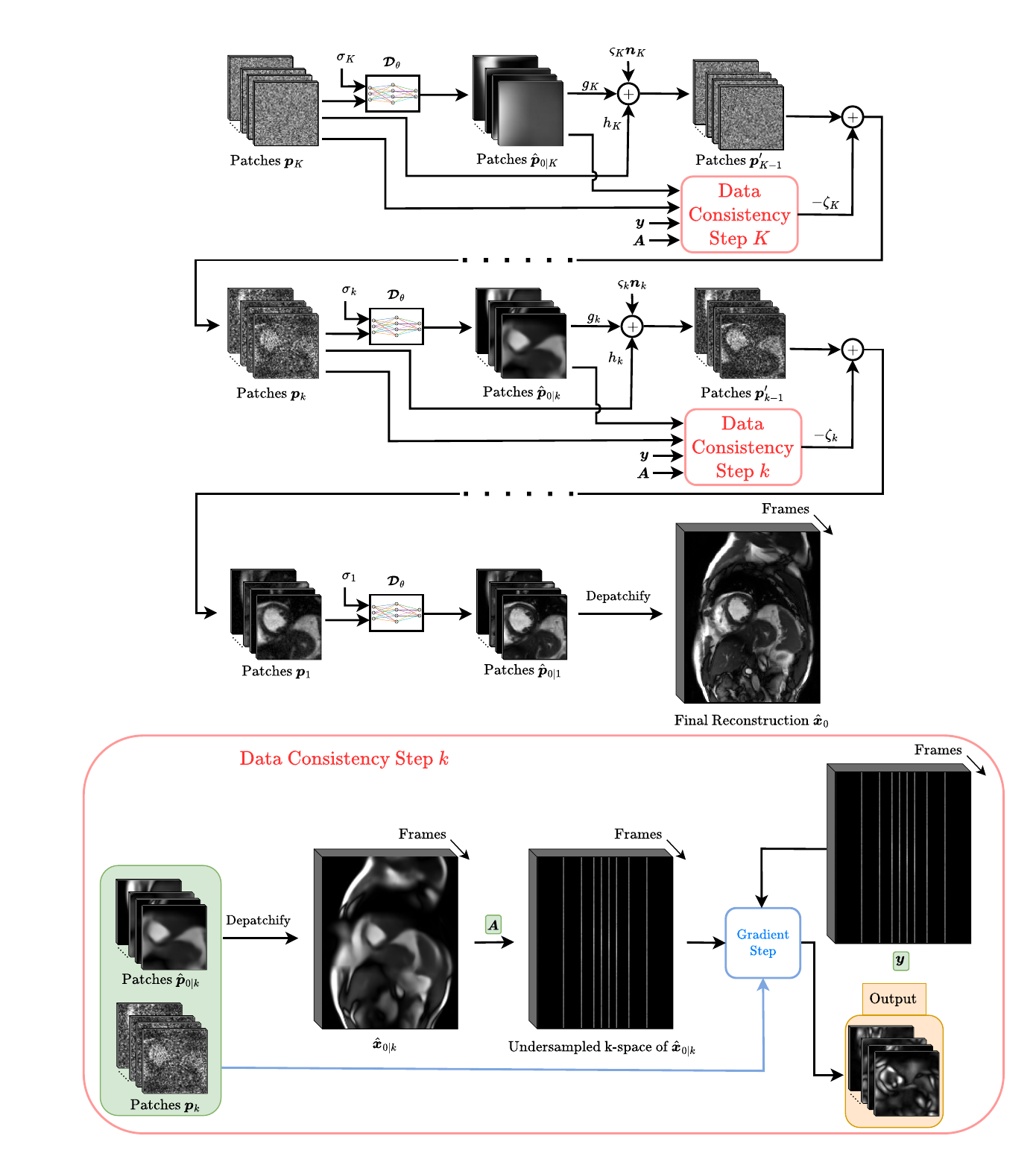}
  \caption{Visual representation of the proposed CineDiff framework.}
  \label{fig:structure}
\end{figure}

In response, we propose a patch-based spatiotemporal diffusion reconstruction framework, called CineDiff, for cardiac cine imaging. CineDiff integrates a learned diffusion prior with the MRI forward model through measurement-guided diffusion sampling. 
After training a patch-based video denoiser $\mathcal{D}_{\vec{\theta}}(\cdot)$, parameterized by $\vec{\theta}$, the inference in CineDiff proceeds as illustrated in Figure~\ref{fig:structure}, with the corresponding sampling procedure summarized in Algorithm~\ref{alg:patch_based_CineDiff}. 

Specifically, CineDiff models the image sequence using overlapping spatiotemporal patches, where $\mathcal{P}(\cdot)$ represents the ``patchification'' operator that maps a cine image series to a set of patches, and $\mathcal{P}^{-1}(\cdot)$ represents the ``de-patchification'' operator that aggregates overlapping patches into a cine image series. To suppress boundary artifacts, we specifically design the operator $\mathcal{P}^{-1}(\cdot)$ to use ramp-weighted averaging in the overlapping regions. Details of these operators are provided in Appendix~\ref{app:patch_operations} in Supplementary Material. The reconstruction in CineDiff is performed via the variance exploding formulation of DDIM~\cite{bendel2025solving} sampling combined with DPS-based data consistency, where each iteration consists of a generative update along the diffusion trajectory followed by data consistency correction using the forward model. At each time step, since data consistency is enforced in the k-space of the entire image series and not patches, the patches are converted to images using $\mathcal{P}^{-1}(\cdot)$ before the DPS-based data consistency step can be applied. To reduce the number of time steps during inference, CineDiff supports a warm-start, where a preliminary reconstruction (e.g., from CS) is used as an initial estimate, which is subsequently refined using diffusion.
%with the reserve diffusion process initialized at an intermediate time step. 

% As mentioned, we use the neural network architecture from the video diffusion model paper~\cite{video_diffusion_pytorch} in the proposed method. The network input size was set to $64\times 64\times 8$ (Height$\times$Width$\times$Time). The number of input channels was set to $2$ to represent the real and imaginary parts of the complex-valued images. The bottleneck dimension was set to 64, and the dimension multiplier was set to $(1, 2, 4, 8)$. The denoiser defined in Equation~(7) of the EDM paper~\cite{edm} was used in the proposed method. The training batch size was 60, the learning rate was set to $2\times 10^{-4}$, and an exponential moving average (EMA) of the model weights was used with the decay set to $0.995$. Training was performed on 16 NVIDIA H100 (Nvidia, Santa Clara, California) GPUs using PyTorch Lightning with distributed data parallel (DDP) and took approximately 7 days. The final model checkpoint was selected based on the Fr\'echet inception distance (FID)~\cite{FID} computed between samples generated unconditionally by the model and the validation dataset.

% The architecture of the denoiser $\mathcal{D}_{\vec{\theta}}(\cdot)$ is based on the video diffusion model described in~\cite{video_diffusion_pytorch}. All cine images were

\subsection{Implementation details of CineDiff}
\subsubsection{Preparation of training data}

The diffusion model was trained on fully sampled cine data from OCMR~\cite{OCMR} and CMRxRecon 2023--2025 repositories~\cite{CMRxRECON}. The summary of training data is provided in \tabref{training}. During preprocessing of the CMRxRecon dataset, we observed temporal inconsistencies in the reconstructed complex-valued cine images, manifested as intermittent sign flips and frame-to-frame fluctuations in image phase and magnitude. These artifacts are consistent with frame-dependent complex scaling ambiguities, where each frame is effectively multiplied by an unknown complex scalar. We believe these preprocessing errors were inadvertently introduced by the original authors of the CMRxRecon work. As a result, some cine series, regardless of the reconstruction method, yielded temporally stable phase but inconsistent magnitude, while others produced stable magnitude with drifting phase. To address this issue, we developed a correction procedure that enforces temporal consistency by aligning global phase across frames and combining complementary reconstructions to retain both stable magnitude and phase. This approach effectively removes the frame-wise ambiguity without requiring access to the original raw data, thereby enabling reliable use of the dataset for training. Finally, the corrected images were normalized on a per-subject basis by dividing by the maximum image magnitude.

\subsubsection{Denoiser training}

% Height and weight of images, as well as the number of frames, in an RT cine series are not fixed and depend on the field-of-view (FOV) and spatial and temporal resolutions. Subjects of different body sizes typically require different FOVs, even when the spatial resolution is kept constant. 

To train the denoiser in CineDiff, we extracted overlapping spatiotemporal patches of size $64\times64\times8$ ($H \times W \times F$) from coil-combined complex-valued images. Rather than sampling patches uniformly over the image, we defined the selection probability using a truncated 2D Gaussian, which biases sampling toward central anatomical structures and away from background regions. The number of input channels was set to $2$ to represent the real and imaginary components of the complex-valued images. Each image series was normalized by dividing by the maximum magnitude within that series.

% The denoiser in Equation~(7) of the EDM paper~\cite{edm} was used, with a bottleneck dimension of 64. 
The base neural network was implemented using the video diffusion network architecture~\cite{video_diffusion_pytorch} with a bottleneck dimension of 64. The denoiser was preconditioned following Equation~7 and trained using the loss function in Equation~104 of the paper by Karras et al.~\cite{edm}.
Training used a batch size of 60, a learning rate of $2\times10^{-4}$, and an exponential moving average with decay 0.995. Training was performed on 16 NVIDIA H100 (Nvidia, Santa Clara, California) GPUs using PyTorch Lightning with distributed data parallel and required approximately 7 days. The final model checkpoint was selected based on the Fr\'echet inception distance~\cite{FID} computed between unconditionally generated samples and the validation dataset.

% The architecture of the denoiser $\mathcal{D}_{\vec{\theta}}(\cdot)$ is based on the video diffusion model described in~\cite{video_diffusion_pytorch}. All cine images were

% However, the input image size for a diffusion model is fixed and cannot be changed once training is complete. \todo{Moreover, large-scale cardiac MRI datasets remain limited, whereas diffusion models are typically trained with large amounts of data~\cite{DDPM,edm}}. To address these issues, we adopted idea of patch-based diffusion models paper~\cite{CT_patch_based_diffusion} and patchified our cine training series of shape Height$\times$Width$\times$Time ($H \times W \times T$) into patches $\vec{p}$ of fixed size 64$\times$64$\times$8 ($H \times W \times T$). This patch-based strategy also requires less training data than training a full-frame diffusion model~\cite{CT_patch_based_diffusion}.

% Rather than extracting these patches uniformly over the pixel dimensions, we apply a truncated Gaussian distribution along the height and width dimensions. This strategy generates more patches from relevant anatomical structures, rather than empty air areas. For the time dimension, we perform uniformly patchifing treated all frames equally.

\subsubsection{Inference}

Since fully sampled data are generally unavailable at the inference stage, the maximum magnitude in an image series was estimated from the inverse Fourier transform of the temporally averaged k-space. For normalization, the original k-space data were divided by this value before reconstruction. After reconstruction, the resulting images were multiplied by the same value to ensure that the reconstruction scale was consistent with the original input.

For each cine series, reconstruction was performed on a single NVIDIA H100 GPU. Using CS reconstructions or another baseline reconstruction as a starting point can reduce reconstruction time in proportion to the number of skipped sampling steps. For Study~I, CS-warm-start allowed reconstruction from 50 steps. For Study~II and Study~III, where CS is less precise, we used 200 and 500 steps, respectively. To improve image SNR, we relied on posterior averaging, i.e., we generated eight ($N_{\sf AVG}=8$) reconstructions using different noise realizations and averaged the resulting complex-valued image series on a pixel-wise basis. With $N_{\sf AVG}=8$, reconstruction took approximately 5 to 8 minutes per series in Study~I. %Without posterior averaging, i.e., $N_{\sf AVG}=1$, the reconstruction time can be reduced by a factor of eight.

% The maximum magnitude of the reconstruction was estimated from the inverse FFT (iFFT) of the temporally averaged k-space. The original k-space data were divided by this value before reconstruction. After reconstruction, the resulting images were multiplied by the same value to ensure that the reconstruction scale was consistent with the original input.

% For each cine series, reconstruction was performed on a single NVIDIA H100 (Nvidia, Santa Clara, California) GPU. Depending on the cine series size, reconstruction took approximately 5 to 8 minutes per series. To improve image SNR, we generated reconstructions eight times using different random seeds and then averaged the results. Using CS reconstructions or another baseline reconstruction as a warm start can reduce reconstruction time in proportion to the number of sampling steps. In this paper, we used CS reconstructions for the warm start. The differences between using and not using a warm start can be found in~\ref{tab:cine_metrics}. A warm start with CS reconstruction was used for prospective studies. For Study~II, we started from the last 200 steps; for Study~III, we started from the last 500 steps.

\subsection{Image analysis}

For fully sampled breath-holding cine data (Study~I), where ground truth was available, we assessed reconstructed cine-series quality using peak signal-to-noise ratio (PSNR), structural similarity index measure (SSIM), learned perceptual image patch similarity (LPIPS)~\cite{LPIPS}, and deep image structure and texture similarity (DISTS)~\cite{DISTS}. 
%\textr{RA: Add citations for LPIPS and DISTS}. 
PSNR was computed on coil-combined complex-valued image series, while SSIM, LPIPS, and DISTS were computed on the coil-combined magnitude images, averaged over frames. All metrics were computed from the central region of the image series to focus on the dynamic cardiac structures and avoid bias from largely static peripheral tissues or the air regions that are not clinically evaluated. 
For free-breathing RT cine data (Studies II and III), ground truth was not available. The results were scored in a blinded manner by two experienced readers, including a CMR-trained cardiologist with 10 years of experience. For each cine series, reconstructions from the three methods (CS, CineVN, and CineDiff) were anonymized and presented side-by-side in randomized order. Each reader assigned an overall image quality score from 1 (worst) to 5 (best) and an image sharpness score from 1 (worst) to 5 (best) for each cine series. Across all studies, CineDiff was compared against CS~\cite{Chen2019SCoRe} and CineVN~\cite{CineVN}.

\subsection{Experiments}

\subsubsection{Study I: Retrospectively undersampled data}
% \textr{retrospectively undersampled healthy subjects, breath-hold}
% \textr{\\3T/1.5T}

Held-out fully sampled breath-hold cine data from healthy volunteers in OCMR~\cite{OCMR} and the CMRxRecon 2024 dataset~\cite{CMRxRECON} were used as test data for performance evaluation. 
From the OCMR dataset, we used 10 slices, covering short-axis (SAX), 2-chamber (2Ch), and 4-chamber (4Ch) views. 
From the CMRxRecon 2024 dataset, we used 20 slices, covering the same cardiac imaging views (SAX, 2Ch, and 4Ch). 
All data were acquired using balanced steady-state free precession (bSSFP) sequences on multiple MR scanners at field strengths of 1.5~T and 3~T (Vida, Sola, Prisma, Avanto, and Cima.X from Siemens Healthineers and UMR670, UMR780, UMR790, and UMR880 from United Imaging Healthcare). Details of the datasets are provided in \tabref{study_i}. 
The fully sampled breath-held data were retrospectively undersampled at acceleration rates of $R \in$ \{8, 12, 16, 20\} using %The GRO sampling pattern was use. Details of the sampling pattern can be found in section \textr{???}.
the variable density golden ratio offset (GRO)~\cite{GRO_sampling_pattern} Cartesian sampling pattern. The coil sensitivity maps were inferred from the time-averaged undersampled k-space data using an eigenvalue approach to autocalibrating parallel MRI (ESPIRiT)~\cite{uecker2014}.

\subsubsection{Study II: Prospectively undersampled real-time data}
% \textr{\\prospectively undersampled healthy subjects, free-breathing}
% \textr{RT, 3T/1.5T}

A total of 15 free-breathing RT cine series from 15 patients were randomly selected from the OCMR dataset, including SAX, 2Ch, and 4Ch views of the heart. The data were acquired using a bSSFP sequence on 1.5T (Sola) and 3T (Vida) scanners. The data were prospectively undersampled at $R=9$ using the GRO Cartesian sampling pattern. The coil sensitivity maps were inferred from the time-averaged undersampled k-space data using ESPIRiT. Additional details of the data used in this study are provided in \tabref{study_ii_iii}. 

\subsubsection{Study III: Prospectively undersampled real-time data at mid-field}
% \textr{\\prospectively undersampled patient/porcine, free-breathing}
% \textr{\\0.55T}
%\textr{out-of-distribution: low field}

A total of 10 prospectively undersampled free-breathing RT cine slices were acquired on a 0.55T system (MAGNETOM Free.Max, Siemens Healthineers). Specifically, 5 slices from healthy volunteers and 5 slices from a porcine study were included, consisting of both SAX and 2Ch views. The data were prospectively undersampled at $R=10$ using the GRO Cartesian sampling pattern, and the coil sensitivity maps were inferred from the time-averaged undersampled k-space data using ESPIRiT. Additional details of the data used in this study are provided in \tabref{study_ii_iii}.

%\subsection{Data processing and diffusion model pre-training}

\section{Results}

\subsection{Study I: Retrospectively undersampled data}
Table~\ref{tab:cine_metrics} summarizes image quality metrics averaged over 30 cine series for $R=8, 12, 16,$ and $20$. We evaluated four configurations of CineDiff that vary in the number of reverse diffusion steps (controlled by the starting timestep $T$), initialization, and posterior averaging ($N_{\sf AVG}$). Here, $T=1{,}000$ denotes reconstruction from Gaussian noise (no warm start), whereas $T=50$ denotes truncated sampling over the final 50 diffusion steps. Since the reconstruction time has a linear dependence on $T$, $T=50$ offers 20 times faster sample generation compared to $T=1{,}000$. Initialization (Init) indicates the use of CS reconstruction as the starting point, and $N_{\sf AVG}$ denotes the number of reconstructions with different random seeds used for posterior averaging.

In almost all cases, CineDiff outperforms CS and CineVN. Across different CineDiff configurations, $T=50$ with CS initialization and $N_{\sf AVG}=8$ yields the highest PSNR and SSIM values, while $T=50$ with CS initialization and $N_{\sf AVG}=1$ yields the lowest LPIPS and DISTS values. Overall, CS initialization improves reconstruction and reduces inference time. Since the posterior averaging can make the images softer-looking, it benefits PSNR and SSIM but degrades the perceptual quality metrics, such as LPIPS and DISTS.

Figure~\ref{fig:retro_P014_cine_sax_sli2_172x246_R_16} and the corresponding Video~S1 in the Supplementary Material show a representative example at $R=16$. Using CS initialization with $N_{\sf AVG}=8$, CineDiff better preserves fine structures and yields sharper reconstructions than CS and CineVN, as indicated by the red arrows. An additional example is provided as Figure~\ref{fig:retro_P014_cine_lax_sli1_172x168_R_16} and the corresponding Video~S5 in the Supplementary Material. % Additional examples at $R=8$, $12$, and $20$ are provided in the Supporting Information (Figs.~S?–S?).

\subsection{Study II: Prospectively undersampled real-time data}
The first column of Table~\ref{tab:reader_study} summarizes reader scores for free-breathing RT cine acquired in patients at 1.5T or 3T. Scores are averaged across 15 cine series and two experienced readers. CineDiff reconstructions were generated using CS initialization with truncated sampling ($T=200$) and $N_{\sf AVG}=8$.

CineDiff achieves the highest scores in both overall image quality and image sharpness compared to CS and CineVN. The improvement is consistent across readers and reflects better preservation of fine structures and reduced blurring artifacts.

Figure~\ref{fig:us0114_f000} and the corresponding Video~S2 in the Supplementary Material show a representative example, where CineDiff produces sharper myocardial borders and improved delineation of anatomical structures relative to CS and CineVN. An additional example is provided as Figure~\ref{fig:us0011_f000} and the corresponding Video~S6 in the Supplementary Material.

% The first column of Table~\ref{tab:reader_study} provides image quality scores from 2 expert readers for the free-breathing RT patient study conducted on the 3T/1.5T scanner. The shown results are averaged across 15 slices. CineDiff results with CS initialization from the final 200 steps were used for scoring. As seen in the table, CineDiff received both the highest image quality score and the highest image sharpness score. 
% A representative result is shown in figure~\ref{fig:pros_RTFB_recon} where CineDiff shows finer details and sharper edges compared to CS and CineVN methods. Additional representative images for study II are shown in Figures S? and S? in the Supporting Information.

\subsection{Study III: Prospectively undersampled real-time data at mid-field}

The second and third columns of Table~\ref{tab:reader_study} summarize reader scores for 0.55T free-breathing RT cine in human and porcine studies, respectively (averaged across 5 slices and two readers). CineDiff reconstructions were generated using $N_{\sf AVG}=8$ and $T=500$.

In the 0.55T human study, CineDiff achieves the highest scores for both overall image quality and image sharpness. In the porcine study, CineDiff matches CineVN in overall image quality while achieving the highest sharpness score. Across both settings, CineDiff consistently improves edge definition and structural clarity relative to CS and remains competitive or superior to CineVN.

Figures~\ref{fig:us9994_f000} and ~\ref{fig:pros_porcine_recon} and the corresponding Videos~S3 and S4 in the Supplementary Material show representative examples, from human and porcine studies, respectively. Consistent with prior studies, CineDiff yields sharper boundaries and improved visualization of anatomical details. Additional examples are provided in the Supplementary Material Figures~\ref{fig:us9993_f000} and \ref{fig:us8997_f000} along with the corresponding Videos~S7 and S8.

% The second and third columns of Table~\ref{tab:reader_study} report reader scores from 2 expert readers for the 0.55~T free-breathing RT human study and the 0.55~T porcine study, respectively (both averaged across 5 slices). In the 0.55~T human study, CineDiff achieved the highest scores for both overall image quality and image sharpness. In the 0.55~T porcine study, CineDiff and CineVN were tied in overall image quality, while CineDiff achieved the highest sharpness score. These results indicate that CineDiff maintains strong performance on out-of-distribution low-field acquisitions, with consistently superior sharpness and competitive or best overall image quality.
% A representative result is shown in figure~\ref{fig:pros_porcine_recon}. As pointed out in the Results section, CineDiff generates images with more defined boundaries. Additional representative images for study III are shown in Figures S? and S? in the Supporting Information.

\section{Discussion}

In this work, we proposed and evaluated CineDiff, a patch-based diffusion reconstruction framework for accelerated 2D real-time cine CMR. Across all settings, including breath-held data with retrospective undersampling (Study I), free-breathing RT data with prospective undersampling in human subjects (Study II), and free-breathing RT data with prospective undersampling in both human and porcine models at lower field (Study III), CineDiff achieved superior or competitive image quality compared with CS and the supervised method CineVN.

Unlike most other diffusion-based image recovery methods, CineDiff employs patch-based training, which ameliorates the issue of limited training data. To avoid the patch-boundary artifacts, CineDiff uses a ramp-weighted de-patchification strategy for merging overlapping patches during inference. Simple averaging introduces visible boundary artifacts at patch edges, which can manifest as structured discontinuities and confound anatomical interpretation. The proposed ramp weighting progressively down-weights contributions near patch boundaries and emphasizes patch centers, effectively suppressing these artifacts and improving spatial consistency.

In the retrospective setting (Study I), CineDiff with CS initialization and $N_{\sf AVG}=8$ achieved the highest PSNR and SSIM scores across all acceleration rates. For example, at $R=16$, CineDiff achieved a PSNR/SSIM of 30.54/0.9149, compared with 29.36/0.9007 for CineVN and 28.67/0.8676 for CS. As a posterior sampling method, CineDiff enables traversal of the perception–distortion tradeoff~\cite{blau2018perception}. Without posterior averaging ($N_{\sf AVG}=1$), reconstructions exhibit higher perceptual realism, as reflected by lower LPIPS and DISTS values. With posterior averaging ($N_{\sf AVG}>1$), pixel-wise variance is reduced, yielding higher PSNR and SSIM at the cost of slightly smoother appearance. For instance, at $R=16$, CineDiff achieved an LPIPS/DISTS of 0.0737/0.0968 for $N_{\sf AVG}=1$, compared with 0.0755/0.1058 for $N_{\sf AVG}=8$. Nonetheless, the reconstruction with $N_{\sf AVG}=8$ is visually sharper and better delineates fine structures, such as papillary muscles, compared to CS and CineVN (Figure~\ref{fig:retro_P014_cine_sax_sli2_172x246_R_16}). Posterior samples can also be used to estimate pixel-wise uncertainty via sample variance~\cite{wen2024task}, although this was not the primary focus of this work.

CS warm-start ($T=50$) further improved performance by both reducing reconstruction time and enhancing reconstruction quality. Relative to Gaussian initialization ($T=1{,}000$), warm-starting reduced the number of reverse steps by a factor of 20 while yielding consistent gains in PSNR and SSIM. This behavior is consistent with prior work on warm-started diffusion~\cite{scholz2025warm}, which shows that initialization from an informed estimate reduces the distance to the target distribution and shortens the denoising trajectory. In inverse problems, where measurements already constrain the solution space, this places the sample closer to the posterior manifold and allows the diffusion process to focus on local refinement. Although more advanced deep-learning-based initializations are possible, we selected CS due to its computational efficiency and broad availability.

In the free-breathing RT studies with prospective undersampling (Studies II–III), where no reference standard was available, CineDiff achieved the highest expert scores for image quality and sharpness. In Study II (1.5T/3T), CineDiff achieved mean scores of 4.97 (quality) and 4.93 (sharpness), compared with 4.27/4.30 for CineVN and 4.30/4.27 for CS. These near-ceiling scores indicate diagnostically acceptable image quality and are consistent with the qualitative results in Figures~\ref{fig:us0114_f000} and \ref{fig:us0011_f000}, where CineDiff shows sharper myocardial boundaries and improved depiction of cardiac structures. Scores were lower for all methods in Study III (0.55T), reflecting the more challenging acquisition conditions. Nonetheless, CineDiff outperformed both CS and CineVN in human subjects and porcine models. In the human example (Figure~\ref{fig:us9994_f000}), CineDiff provides a clearer blood–myocardium boundary, whereas CineVN and CS exhibit blocky artifacts. A similar trend is observed in the porcine example (Figure~\ref{fig:pros_porcine_recon}), where CineVN and CS produce blurred images with blocky artifacts, while CineDiff maintains sharper and more consistent boundary definition.

The strong generalization of CineDiff to mid-field acquisitions and across species is notable, given that the model was trained exclusively on 1.5T/3T breath-held cine data. This generalization likely arises from two factors. First, the patch-based training strategy emphasizes local spatiotemporal features—such as myocardial boundaries, blood-pool interfaces, and trabecular structures—that are relatively consistent across field strengths and species. Second, the DPS data-consistency step enforces agreement with measured k-space data at each sampling iteration, which constrains the reconstruction to remain faithful to the acquisition and mitigates the risk of hallucinated features.

Finally, the choice of diffusion length was adapted to the quality of the CS warm start. While $T=50$ was sufficient for Study I, higher values ($T=200$ and $T=500$) were required in Studies II and III, respectively. This reflects the reduced quality of CS initialization in RT cine due to respiratory motion and, in Study III, a lower signal-to-noise ratio. Starting the reverse process at a higher noise level reduces reliance on imperfect initialization and allows stronger refinement by the diffusion prior. %In contrast, using $T=50$ in these settings led to occasional residual artifacts.

\section{Limitations}
This study has several limitations that should be considered when interpreting the results. First, the prospective evaluation in Studies II and III is based on a relatively small number of slices, which may limit statistical power and restrict the generalizability of the findings. Larger and more diverse datasets, ideally across multiple centers, are needed to confirm robustness across different patient populations and acquisition settings. Second, CineDiff remains computationally demanding. Although CS warm-start substantially reduces the number of diffusion steps, reconstruction with posterior averaging still requires several minutes per cine series on a high-end GPU. %Posterior averaging further improves PSNR and SSIM but increases wall-clock time proportionally, which may hinder clinical translation in resource-constrained settings. 
Third, the evaluation focuses on image-quality metrics and blinded reader scores, which, while standard in reconstruction studies, do not directly assess clinical utility. Additional validation on downstream functional endpoints, such as ventricular volumes, ejection fraction, and regional wall motion, is needed to establish clinical relevance. Finally, several design choices, including patch size, overlap, and sampling parameters, were selected empirically rather than through systematic optimization. These factors may influence the trade-off between local detail preservation and global consistency, and further tuning may yield additional performance improvements. 

\section{Conclusions}
We presented and evaluated CineDiff, a patch-based diffusion reconstruction framework for highly accelerated 2D RT cine CMR. The method incorporates two key design choices: training on complex-valued spatiotemporal patches and reconstruction using ramp-weighted de-patchification to reduce boundary artifacts. Across both retrospective and prospective undersampled studies, CineDiff consistently improved image quality and sharpness at high acceleration rates, including in mid-field and porcine acquisitions. These results support CineDiff as a robust reconstruction approach for RT cine imaging in challenging acquisition settings.

\newpage
% \appendix
% \section*{Code availability}
% The code will be made available after the manuscript has been accepted for publication.

\section*{Acknowledgments and funding}
This work was funded by NIH grants R01-EB029957 and R01-HL151697. We thank Dr. Yuchi Han for scoring images and providing feedback.

\section*{Ethics Declarations}
\begin{itemize}
% \item Funding
\item \textbf{Declaration of competing interests}: The authors declare no competing interests.
\item \textbf{Ethics approval and consent}: For the human subject data, approval was granted by the Institutional Review Board (IRB) at The Ohio State University (2020H0402 and 2019H0076). Informed consent to participate in the study and publish results was obtained from all individual participants.
\item \textbf{Data and code availability}: Code and a sample dataset will be made publicly available after the peer-review process is complete.
\item \textbf{Author contributions}: X. Lei implemented CineDiff and generated results. P. Schniter assisted with CineDiff optimization and co-mentored the first author. J. Varghese assisted with the study design and image analysis. R. Ahmad supervised all aspects of the study and co-mentored the first author.
\end{itemize}

\printbibliography

% \include{figures/cache/MF_table}

% We have presented and evaluated CineDiff, a patch-based diffusion reconstruction framework for highly accelerated 2D real-time cine CMR. Two key features of CineDiff are that the diffusion model was trained on complex-valued cine-series patches and overlapping patches were used during reconstruction. Retrospective and prospective experiments demonstrate that CineDiff achieves better overall image quality and better sharpness from highly undersampled data, including low-field and porcine acquisitions.
% Table for the diffusion training
\clearpage
\begin{table}[ht]%[h]
\centering
\renewcommand{\arraystretch}{1.15}
\setlength{\tabcolsep}{8pt}
\begin{tabular}{llcc}
\toprule
% \rowcolor{gray!25}
& & \textbf{Train} & \textbf{Validation} \\
\midrule

% testing_max1_ocmr232425_CS.pkl:length=517
% testing_max1_ocmr232425_CS_all.pkl:length=1245
% testing_station6_max1_ocmr232425_CS.pkl:length=618
% testing_station7_max1_ocmr232425_CS.pkl:length=110
% training_max1_ocmr232425_CS.pkl:length=4712
% training_max1_ocmr232425_CS_first1k.pkl:length=100
% 94+108+180+113 & 10+12+20+12+94
% \multicolumn{2}{l}{\textbf{Total (subjects)}} & 495 & 148  \\

\multicolumn{2}{l}{Total cine series}   & 4,712 & 1,245  \\

\multicolumn{2}{l}{Field strength} & \multicolumn{2}{c}{1.5T and 3T} \\

% \rowcolor{gray!10}
\multicolumn{2}{l}{Orientation} & \multicolumn{2}{c}{SAX, 2Ch and 4Ch} \\

%OCMR
% Temporal phases      : 10-31
% Temporal resolution  : 27.50-54.72 ms
% Spatial resolution   : 1.172-2.750 mm
% Slice thickness      : 6.000-8.000 mm

%xrecon 2023 
%info is not provided by the data publisher

%xrecon 2024
% Temporal phases     : 14-43
% TR(ms)              : 39.240-47.880
% Spatial resolution  : 1.317-1.861 mm
% Slice thickness     : 6.000-8.000 mm

%Xrecon2025
% Temporal phases      : 12-30
% TR(ms), filtered     : 34.000-46.480
% Spatial resolution   : 0.909-2.262 mm
% Slice thickness      : 6.000-8.000 mm

Frames               & \multicolumn{3}{c}{10--43} \\
Temporal resolution  & \multicolumn{3}{c}{27.50--54.72 ms} \\
Spatial resolution   & \multicolumn{3}{c}{0.91--2.75 mm} \\
Slice thickness      & \multicolumn{3}{c}{6--8 mm} \\
\bottomrule
\end{tabular}
\caption{Summary of fully sampled training and validation folds used to train the diffusion model. The validation fold was used for hyperparameter tuning and checkpoint selection.}
\label{tab:training}
\end{table}

% Table for the retrospective validation/testing datasets
\clearpage
\begin{table}[ht]%[h]
\centering
\renewcommand{\arraystretch}{1.15}
\setlength{\tabcolsep}{8pt}
\begin{tabular}{llcc}
\toprule
% \rowcolor{gray!25}
& & \textbf{Validation} & \textbf{Test} \\
\midrule

% \multicolumn{2}{l}{\textbf{Total (subjects)}} & 8 &  19 \\
% \rowcolor{gray!10}
\multicolumn{2}{l}{Total cine series}   & 8 & 30  \\

\multicolumn{2}{l}{Acceleration} & \multicolumn{2}{c}{8, 12, 16, 20} \\

\multicolumn{2}{l}{Field strength} & \multicolumn{2}{c}{1.5T and 3T} \\

% \rowcolor{gray!10}
\multicolumn{2}{l}{Orientation} & \multicolumn{2}{c}{SAX, 2Ch and 4Ch} \\

Frames              & & 21--27          & 10--31 \\
Temporal resolution & & 34.20--47.88 ms & 30.80--54.72 ms \\
Spatial resolution  & & 1.34--2.08 mm & 1.32--2.08 mm \\
Slice thickness     & & 6-8 mm & 6--8 mm \\
\bottomrule
\end{tabular}
\caption{Summary of fully sampled validation and testing folds used to assess reconstruction quality from retrospectively undersampled data in Study I. The validation fold was used for hyperparameter tuning.}
\label{tab:study_i}
\end{table}

% Table for the prospective blinded score datasets
\clearpage
\begin{table}[ht]%[t]
\centering
\renewcommand{\arraystretch}{1.15}
\setlength{\tabcolsep}{8pt}
\begin{tabular}{llccc}
\toprule
% \rowcolor{gray!25}
& & \textbf{RT (human)} & \textbf{RT (human)} & \textbf{RT (porcine)} \\
\midrule

% \multicolumn{2}{l}{\textbf{Total cine series}} & 15 &  2 & 2 \\
% \rowcolor{gray!10}
\multicolumn{2}{l}{Total cine series}   & 15 &  5 & 5  \\

\multicolumn{2}{l}{Acceleration} & 9 & 10 & 10 \\

\multicolumn{2}{l}{Field strength} & 1.5T and 3T & 0.55T & 0.55T \\

% \rowcolor{gray!10}
\multicolumn{2}{l}{Orientation} & SAX, 2Ch and 4Ch & 2Ch, 3Ch and 4Ch & SAX, 2Ch and 4Ch \\

%blinded score
%OCMR RT-FB patient 1.5T/3T
% Temporal phases      : 64-151
% Temporal resolution  : 32.90-45.30 ms
% Spatial resolution   : 1.719-2.481 mm
% Slice thickness      : 8.000-8.000 mm

%RT-FB patient 0.55T
% Temporal phases      : 34-35
% Temporal resolution  : 42.00-44.00 ms
% Spatial resolution   : 1.823-3.375 mm
% Slice thickness      : 8.000-8.000 mm

%RT-FB porcine 0.55T
% Temporal phases      : 30-34
% Temporal resolution  : 44.00-49.30 ms
% Spatial resolution   : 1.667-2.625 mm
% Slice thickness      : 8.000-8.000 mm

Frames              & & 64--151          & 34--35          & 30--34 \\
Temporal resolution & & 32.90--45.30 ms  & 42.00--44.00 ms  & 44.00--49.30 ms \\
Spatial resolution  & & 1.719--2.48 mm  & 1.82--3.38 mm & 1.67--2.62 mm \\
Slice thickness     & & 8 mm        & 8 mm        & 8 mm \\
\bottomrule
\end{tabular}
\caption{Summary of free-breathing prospectively undersampled data used in Studies II and III.}
\label{tab:study_ii_iii}
\end{table}

\sisetup{
  mode = text,
  group-digits = false,
  round-mode = places,
  detect-weight = true,
  detect-family = true,
  separate-uncertainty = true
}

\begin{table}[ht]
  \centering
  \small
  \setlength{\tabcolsep}{4.5pt}
  \begin{tabular}{
    l
    c
    c
    c
    c
    S[
      table-format=2.2(2),
      round-precision=2
    ] % PSNR; enter 32.07 or 33.49(39)
    S[
      table-format=1.4(4),
      round-precision=4
    ] % SSIM
    S[
      table-format=1.4(4),
      round-precision=4
    ] % LPIPS
    S[
      table-format=1.4(4),
      round-precision=4
    ] % DISTS
  }
    \toprule
    Method & $R$ & $T$ & $\vec{x}_{\sf init}$ & $N_{\sf AVG}$ & {PSNR $\uparrow$} & {SSIM $\uparrow$} & {LPIPS $\downarrow$} & {DISTS $\downarrow$} \\
    \midrule

    % ----------------- R = 8 -----------------
    CineDiff & \multirow{6}{*}{8}  & 50   & CS            & 1            & 32.10(36)           & 0.9375(39)           & \bfseries 0.0486(36) & \bfseries 0.0760(34) \\
    CineDiff &                     & 50   & CS            & 8            & \bfseries 33.56(38) & \bfseries 0.9486(35) & 0.0528(41)           & 0.0878(36) \\
    CineDiff &                     & 1000 & \textemdash   & 1            & 32.05(39)           & 0.9360(41)           & 0.0493(38)           & \bfseries 0.0760(36) \\
    CineDiff &                     & 1000 & \textemdash   & 8            & 33.50(39)           & 0.9477(37)           & 0.0530(42)           & 0.0879(38) \\
    CineVN   &                     & \textemdash & \textemdash & \textemdash & 32.61(44)           & 0.9477(39)           & 0.0543(42)           & 0.0818(46) \\
    CS       &                     & \textemdash & \textemdash & \textemdash & 31.97(44)           & 0.9295(51)           & 0.0759(56)           & 0.1044(52) \\
    \midrule

    % ----------------- R = 12 ----------------
    CineDiff & \multirow{6}{*}{12} & 50   & CS            & 1            & 30.38(39)           & 0.9155(50)           & \bfseries 0.0622(42) & \bfseries 0.0863(38) \\
    CineDiff &                     & 50   & CS            & 8            & \bfseries 31.83(40) & \bfseries 0.9317(44) & 0.0649(45)           & 0.0978(37) \\
    CineDiff &                     & 1000 & \textemdash   & 1            & 30.26(40)           & 0.9129(52)           & 0.0628(41)           & 0.0865(37) \\
    CineDiff &                     & 1000 & \textemdash   & 8            & 31.74(40)           & 0.9302(45)           & 0.0651(46)           & 0.0974(39) \\
    CineVN   &                     & \textemdash & \textemdash & \textemdash & 30.51(46)           & 0.9220(58)           & 0.0749(57)           & 0.0976(55) \\
    CS       &                     & \textemdash & \textemdash & \textemdash & 30.07(44)           & 0.8992(66)           & 0.1033(68)           & 0.1283(58) \\
    \midrule

    % ----------------- R = 16 ----------------
    CineDiff & \multirow{6}{*}{16} & 50   & CS            & 1            & 29.13(42)           & 0.8949(66)           & \bfseries 0.0737(49) & \bfseries 0.0968(44) \\
    CineDiff &                     & 50   & CS            & 8            & \bfseries 30.54(43) & \bfseries 0.9149(57) & 0.0755(50)           & 0.1058(39) \\
    CineDiff &                     & 1000 & \textemdash   & 1            & 28.93(42)           & 0.8912(68)           & 0.0761(50)           & 0.0983(45) \\
    CineDiff &                     & 1000 & \textemdash   & 8            & 30.50(42)           & 0.9142(56)           & 0.0754(50)           & 0.1043(40) \\
    CineVN   &                     & \textemdash & \textemdash & \textemdash & 29.36(48)           & 0.9007(73)           & 0.0915(66)           & 0.1131(62) \\
    CS       &                     & \textemdash & \textemdash & \textemdash & 28.67(46)           & 0.8676(86)           & 0.1267(76)           & 0.1456(62) \\
    \midrule

    % ----------------- R = 20 ----------------
    CineDiff & \multirow{6}{*}{20} & 50   & CS            & 1            & 28.21(44)           & 0.8763(82)           & 0.0870(60)           & \bfseries 0.1069(49) \\
    CineDiff &                     & 50   & CS            & 8            & \bfseries 29.61(45) & \bfseries 0.9010(69) & \bfseries 0.0854(57)           & 0.1138(45) \\
    CineDiff &                     & 1000 & \textemdash   & 1            & 27.79(40)           & 0.8699(79)           & 0.0901(61)           & 0.1092(48) \\
    CineDiff &                     & 1000 & \textemdash   & 8            & 29.55(44)           & 0.9006(68)           & \bfseries 0.0854(57) & 0.1136(46) \\
    CineVN   &                     & \textemdash & \textemdash & \textemdash & 28.23(48)           & 0.8781(92)           & 0.1151(83)           & 0.1312(76) \\
    CS       &                     & \textemdash & \textemdash & \textemdash & 27.80(47)           & 0.8461(98)           & 0.1447(86)           & 0.1579(68) \\
    \bottomrule
  \end{tabular}
  \caption{Quantitative cine CMR reconstruction results (mean $\pm$ standard error) at accelerations $R\in\{8,12,16,20\}$. $T$ is the total number of denoising steps. $N_{\sf AVG}$ denotes the number of posterior samples averaged. Best per $R$ is \textbf{bold}.}
  \label{tab:cine_metrics}
\end{table}

%old one, scores by one expert
% \begin{table}[t]
%   \centering
%   \small
%   \setlength{\tabcolsep}{6pt}
%   \begin{tabular}{lccc}
%     \toprule
%     \multirow{2}{*}{Method} & 3T/1.5T OCMR (human) & 0.55T (human) & 0.55T (porcine) \\
%     & $R{=}9$, $n{=}15$ slices & $R{=}10$, $n{=}5$ slices & $R{=}10$, $n{=}5$ slices \\
%     & $T{=}200$                & $T{=}500$                 & $T{=}500$                 \\
%     \midrule
%     \multicolumn{4}{c}{\textbf{Overall image quality}} \\
%     CineDiff & {\bfseries $\num{4.93} \pm \num{0.07}$} & {\bfseries $\num{3.60} \pm \num{0.24}$} & {\bfseries $\num{3.80} \pm \num{0.37}$} \\
%     CineVN   & $\num{4.20} \pm \num{0.17}$ & $\num{2.80} \pm \num{0.49}$ & {\bfseries $\num{3.80} \pm \num{0.37}$} \\
%     CS       & $\num{4.20} \pm \num{0.20}$ & $\num{2.20} \pm \num{0.37}$ & $\num{2.60} \pm \num{0.24}$ \\
%     \midrule
%     \multicolumn{4}{c}{\textbf{Image sharpness}} \\
%     CineDiff & {\bfseries $\num{4.87} \pm \num{0.09}$} & {\bfseries $\num{3.60} \pm \num{0.24}$} & {\bfseries $\num{3.80} \pm \num{0.20}$} \\
%     CineVN   & $\num{4.13} \pm \num{0.17}$ & $\num{3.20} \pm \num{0.37}$ & $\num{3.60} \pm \num{0.24}$ \\
%     CS       & $\num{4.00} \pm \num{0.14}$ & $\num{2.00} \pm \num{0.32}$ & $\num{2.60} \pm \num{0.24}$ \\
%     \bottomrule
%   \end{tabular}
%   \caption{Blinded reader study results (mean $\pm$ SE) for prospectively undersampled real-time cine data.}
%   \label{tab:reader_study}
% \end{table}

% scores from two experts
\clearpage
\begin{table}[ht]
  \centering
  \small
  \setlength{\tabcolsep}{6pt}
  \begin{tabular}{lccc}
    \toprule
    \multirow{2}{*}{Method} & Study II & \multicolumn{2}{c}{Study III} \\
    & 1.5T/3T (human) & 0.55T (human) & 0.55T (porcine) \\
    & $R{=}9$, $n{=}15$ & $R{=}10$, $n{=}5$ & $R{=}10$, $n{=}5$ \\
    & $T{=}200$                & $T{=}500$                 & $T{=}500$                 \\
    \midrule
    \multicolumn{4}{c}{Overall image quality} \\
    CineDiff & {\bfseries $\num{4.97} \pm \num{0.03}$} & {\bfseries $\num{3.70} \pm \num{0.15}$} & {\bfseries $\num{3.90} \pm \num{0.23}$} \\
    CineVN   & $\num{4.27} \pm \num{0.12}$ & $\num{3.20} \pm \num{0.29}$ & $\num{3.80} \pm \num{0.25}$ \\
    CS       & $\num{4.30} \pm \num{0.12}$ & $\num{2.80} \pm \num{0.29}$ & $\num{3.20} \pm \num{0.25}$ \\
    \midrule
    \multicolumn{4}{c}{Image sharpness} \\
    CineDiff & {\bfseries $\num{4.93} \pm \num{0.05}$} & {\bfseries $\num{3.90} \pm \num{0.18}$} & {\bfseries $\num{4.10} \pm \num{0.18}$} \\
    CineVN   & $\num{4.30} \pm \num{0.12}$ & $\num{3.30} \pm \num{0.21}$ & $\num{3.80} \pm \num{0.20}$ \\
    CS       & $\num{4.27} \pm \num{0.12}$ & $\num{2.60} \pm \num{0.27}$ & $\num{3.00} \pm \num{0.26}$ \\
    \bottomrule
  \end{tabular}
  \caption{Blinded reader study results (mean $\pm$ standard error) for prospectively undersampled RT cine data. CS reconstruction was used as $\vec{x}_{\sf init}$ to guide the diffusion process.}
  \label{tab:reader_study}
\end{table}

% =========================
% Arrow settings for this figure only
% =========================
% ---------- Arrow 1 ----------
\def\ax{33}
\def\ay{68}
\def\adx{1}
\def\ady{1}
\def\alen{4}

% ---------- Arrow 2 ----------
\def\bx{26}
\def\by{42}
\def\bdx{1}
\def\bdy{-0.5}
\def\blen{5}

% ---------- Combine arrows ----------
\def\figarrows{%
    \myarrow{\ax}{\ay}{\adx}{\ady}{\alen}%
    \myarrow{\bx}{\by}{\bdx}{\bdy}{\blen}%
}

% =========================
% Arrow settings for this figure only
% =========================
% ---------- Arrow 1' ----------
\def\axp{38}
\def\ayp{40}
\def\adxp{-1}
\def\adyp{-1}
\def\alenp{7}

% ---------- Arrow 2' ----------
\def\bxp{48}
\def\byp{20}
\def\bdxp{1}
\def\bdyp{0}
\def\blenp{10}

% ---------- Combine arrows ----------
\def\figarrowsyt{%
    \myarrow{\axp}{\ayp}{\adxp}{\adyp}{\alenp}%
    \myarrow{\bxp}{\byp}{\bdxp}{\bdyp}{\blenp}%
}

% =========================
% Figure
% =========================
\begin{figure}[t]
    \centering
    %------------ Row 1: images with titles on top ------------
    \begin{subfigure}[b]{0.23\textwidth}
        \centering
        \subcaption*{\large \sffamily Reference}
        \figuremetrics{PSNR}{SSIM}{LPIPS}{DISTS}
        \begin{overpic}[width=\linewidth]{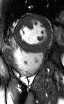}
            \figarrows
        \put(0,68){%
        \tikz{\draw[red!70,dashed,line width=1pt] (0,0) -- (3.79,0);}
        }
        \end{overpic}
    \end{subfigure}
    \hfill
    \begin{subfigure}[b]{0.23\textwidth}
        \centering
        \subcaption*{\large \sffamily CS}
        \figuremetrics{25.92}{0.821}{0.1578}{0.1861}
        \begin{overpic}[width=\linewidth]{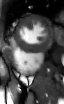}
            \figarrows
        \put(0,68){%
        \tikz{\draw[red!70,dashed,line width=1pt] (0,0) -- (3.79,0);}
        }
        \end{overpic}
    \end{subfigure}
    \hfill
    \begin{subfigure}[b]{0.23\textwidth}
        \centering
        \subcaption*{\large \sffamily CineVN}
        \figuremetrics{26.49}{0.869}{0.1048}{0.1439}
        \begin{overpic}[width=\linewidth]{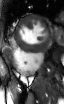}
            \figarrows
        \put(0,68){%
        \tikz{\draw[red!70,dashed,line width=1pt] (0,0) -- (3.79,0);}
        }
        \end{overpic}
    \end{subfigure}
    \hfill
    \begin{subfigure}[b]{0.23\textwidth}
        \centering
        \subcaption*{\large \sffamily CineDiff}
        \figuremetrics{28.55}{0.905}{0.0753}{0.1205}
        \begin{overpic}[width=\linewidth]{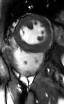}
            \figarrows
        \put(0,68){%
        \tikz{\draw[red!70,dashed,line width=1pt] (0,0) -- (3.79,0);}
        }
        \end{overpic}
    \end{subfigure}

    \vspace{0.5em}
    
    %------------ Row 2: error maps, no titles; ×5 on top-left ------------
    \begin{subfigure}[b]{0.23\textwidth}
        \centering
        \begin{overpic}[width=\linewidth]{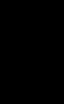}
        \end{overpic}
    \end{subfigure}
    \hfill
    \begin{subfigure}[b]{0.23\textwidth}
        \centering
        \begin{overpic}[width=\linewidth]{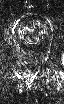}
            \put(0.5,93){\errlabel}
        \end{overpic}
    \end{subfigure}
    \hfill
    \begin{subfigure}[b]{0.23\textwidth}
        \centering
        \begin{overpic}[width=\linewidth]{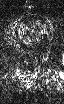}
            \put(0.5,93){\errlabel}
        \end{overpic}
    \end{subfigure}
    \hfill
    \begin{subfigure}[b]{0.23\textwidth}
        \centering
        \begin{overpic}[width=\linewidth]{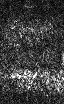}
            \put(0.5,93){\errlabel}
        \end{overpic}
    \end{subfigure}

    \vspace{0.5em}
    %------------ Row 3: time profile, no titles ------------
    \begin{subfigure}[b]{0.23\textwidth}
        \centering
        \begin{overpic}[width=\linewidth,height=2cm]{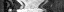}
        \figarrowsyt
        \end{overpic}
    \end{subfigure}
    \hfill
    \begin{subfigure}[b]{0.23\textwidth}
        \centering
        \begin{overpic}[width=\linewidth,height=2cm]{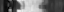}
        \figarrowsyt
        \end{overpic}
    \end{subfigure}
    \hfill
    \begin{subfigure}[b]{0.23\textwidth}
        \centering
        \begin{overpic}[width=\linewidth,height=2cm]{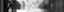}
        \figarrowsyt
        \end{overpic}
    \end{subfigure}
    \hfill
    \begin{subfigure}[b]{0.23\textwidth}
        \centering
        \begin{overpic}[width=\linewidth,height=2cm]{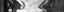}
        \figarrowsyt
        \end{overpic}
    \end{subfigure}
    
    \includegraphics[width=\textwidth]{figures/colorbar_gray.png}

    \caption{Reference image, reconstructions at $R=16$, and corresponding error maps (middle row) and time profiles (bottom row) along the dashed red line. The red arrows highlight details that are better preserved in CineDiff.}
        % \caption{\\Prospective Study\\Reference image, reconstructions at $R=16$,\\and corresponding reconstruction results. The red arrows highlight details that are better preserved in MeanFlow.}
    \label{fig:retro_P014_cine_sax_sli2_172x246_R_16}
\end{figure} % good main

\ReconTriplet
    {us_0114}{000}
    {CS, CineVN, and CineDiff reconstructions from a free-breathing RT prospectively undersampled dataset, along with the time profiles (bottom row) along the dashed red line. The red arrows highlight details that are better preserved in CineDiff.}
    % {Comparison of CS, CineVN, and CineDiff reconstructions for \texttt{us\_0114}, frame 000.}
    {fig:us0114_f000}
    {trim=30 10 20 0} %L, D, R, U
    {
        \myarrow{34}{48}{-1}{0.5}{6}
        \myarrow{43}{30}{-1}{-0.3}{6}
    }
    {{us_0114_1_5T_sli0_100x140_sli0_R_9}
    {59}}
    {
        \myarrow{25}{12}{0.5}{1}{7.6}
        \myarrow{37}{44}{0.5}{1}{7.6}
    }
    {31}
% \ReconTriplet
%     {us_0011}{000}
%     {CS, CineVN, and CineDiff reconstructions from a free-breathing RT prospectively undersampled dataset. The red arrows highlight details that are better preserved in CineDiff.}
%     % {Comparison of CS, CineVN, and CineDiff reconstructions for \texttt{us\_0011}, frame 000. RA: Move the arrows closer to the details.}
%     {fig:us0011_f000}
%     {trim=20 28 0 0} %L, D, R, U
%     {
%         \myarrow{44}{46}{1}{0}{6}
%         \myarrow{28}{57}{1}{-1}{4}
%     }
%     {{us_0011_3T_sli0_100x144_sli0_R_9}
%     {53}}
%     {
%         \myarrow{62}{8}{1}{1}{5.6}
%         \myarrow{36}{34}{1}{-1}{5.6}
%     }
%     {49}

\ReconTriplet
    {us_9994}{000}
    {CS, CineVN, and CineDiff reconstructions from a free-breathing RT prospectively collected from a human subject on a 0.55T scanner, along with the time profiles (bottom row) along the dashed red line. The red arrows highlight details that are better preserved in CineDiff.}
    {fig:us9994_f000}
    {trim=38 0 25 10} %L, D, R, U
    {
        \myarrow{20}{27}{-1}{-0.3}{5.2}
        \myarrow{38}{20}{1}{-1}{4}
    }
    {{us_9994_standard_Free_100x100_sli0_R_10}
    {79}}
    {
        \myarrow{34}{60}{-1}{-0.5}{7.6}
        \myarrow{50}{55}{1}{-0.5}{7.6}
    }
    {23}

% =========================
% Arrow settings for this figure only
% =========================
% ---------- Arrow 1 ----------
\def\ax{51}
\def\ay{70}
\def\adx{-0.5}
\def\ady{1}
\def\alen{6}

% ---------- Arrow 2 ----------
\def\bx{55}
\def\by{38}
\def\bdx{-1}
\def\bdy{-1}
\def\blen{5}

% ---------- Combine arrows ----------
\def\figarrows{%
    \myarrow{\ax}{\ay}{\adx}{\ady}{\alen}%
    \myarrow{\bx}{\by}{\bdx}{\bdy}{\blen}%
}

% =========================
% Arrow settings for this figure only
% =========================
% ---------- Arrow 1' ----------
\def\axp{48}
\def\ayp{45}
\def\adxp{-1}
\def\adyp{-0.5}
\def\alenp{7}

% ---------- Arrow 2' ----------
\def\bxp{58}
\def\byp{10}
\def\bdxp{-1}
\def\bdyp{0.5}
\def\blenp{7}

% ---------- Combine arrows ----------
\def\figarrowsyt{%
    \myarrow{\axp}{\ayp}{\adxp}{\adyp}{\alenp}%
    \myarrow{\bxp}{\byp}{\bdxp}{\bdyp}{\blenp}%
}

\begin{figure}[t]
    \centering
    %------------ Row 1: reconstructions with titles on top ------------
    \begin{subfigure}[b]{0.32\textwidth}
        \centering
        \subcaption*{\large \sffamily CS}
        \begin{overpic}[width=\linewidth, trim=5 90 165 80,clip]{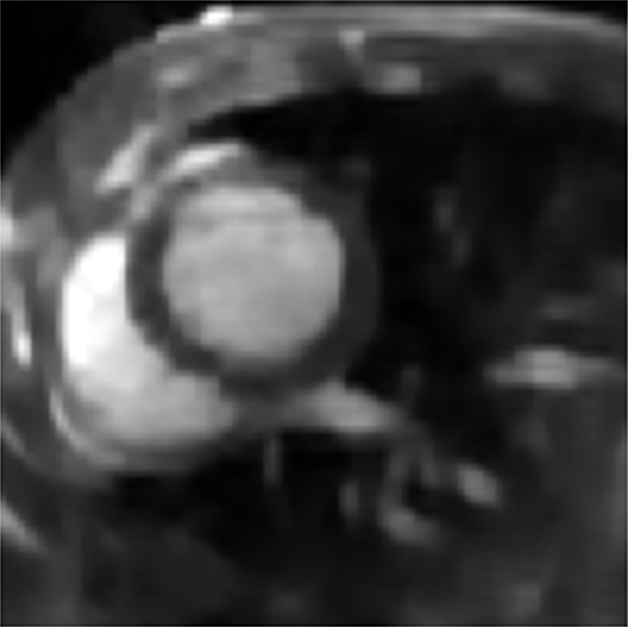}
            \figarrows
        \put(0,57){%
        \tikz{\draw[red!70,dashed,line width=1pt] (0,0) -- (5.2,0);}
        }
        \end{overpic}
    \end{subfigure}
    \hfill
    \begin{subfigure}[b]{0.32\textwidth}
        \centering
        \subcaption*{\large \sffamily CineVN}
        \begin{overpic}[width=\linewidth,trim=5 90 165 80,clip]{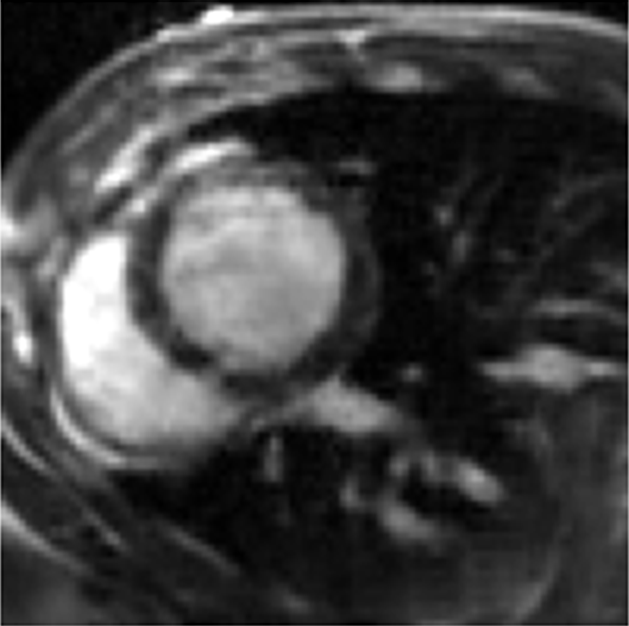}
            \figarrows
        \put(0,57){%
        \tikz{\draw[red!70,dashed,line width=1pt] (0,0) -- (5.2,0);}
        }
        \end{overpic}
    \end{subfigure}
    \hfill
    \begin{subfigure}[b]{0.32\textwidth}
        \centering
        \subcaption*{\large \sffamily CineDiff}
        \begin{overpic}[width=\linewidth,trim=5 90 173 80,clip]{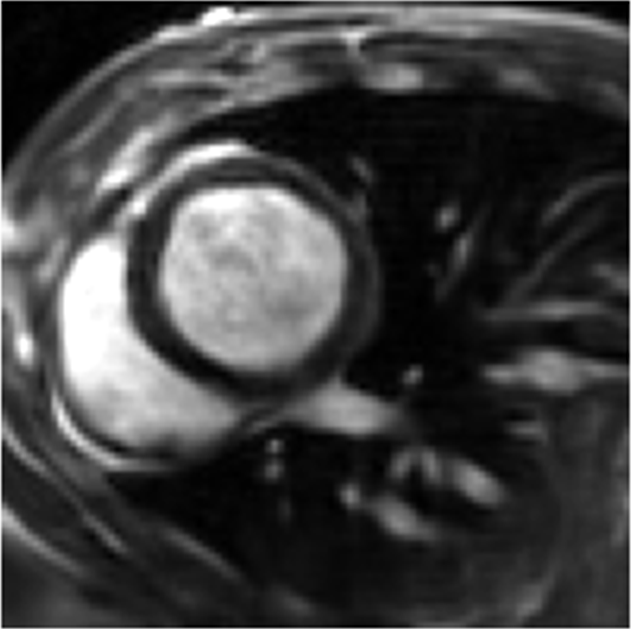}
            \figarrows
        \put(0,57){%
        \tikz{\draw[red!70,dashed,line width=1pt] (0,0) -- (5.2,0);}
        }
        \end{overpic}
    \end{subfigure}

    \vspace{0.5em}
    %------------ Row 3: time profile, no titles ------------
    \begin{subfigure}[b]{0.32\textwidth}
        \centering
        \begin{overpic}[width=\linewidth]{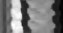}
        \figarrowsyt
        \end{overpic}
    \end{subfigure}
    \hfill
    \begin{subfigure}[b]{0.32\textwidth}
        \centering
        \begin{overpic}[width=\linewidth]{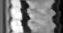}
        \figarrowsyt
        \end{overpic}
    \end{subfigure}
    \hfill
    \begin{subfigure}[b]{0.32\textwidth}
        \centering
        \begin{overpic}[width=\linewidth]{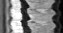}
        \figarrowsyt
        \end{overpic}
    \end{subfigure}

    \includegraphics[width=\textwidth]{figures/colorbar_gray.png}

    \caption{CS, CineVN, and CineDiff reconstructions from a free-breathing RT prospectively collected from a porcine model on a 0.55T scanner, along with the time profiles (bottom row) along the dashed red line. The red arrows highlight details that are better preserved in CineDiff.}
    \label{fig:pros_porcine_recon}
\end{figure}

% \ReconTriplet
%     {us_9978}{000}
%     {CS, CineVN, and CineDiff reconstructions from a free-breathing RT prospectively collected from a human subject on a 0.55T scanner. The red arrows highlight details that are better preserved in CineDiff.}
%     % {Comparison of CS, CineVN, and CineDiff reconstructions for \texttt{us\_9978}, frame 000.}
%     {fig:us9978_f000}
%     {trim=15 35 10 10} %L, D, R, U
%     {
%         \myarrow{22}{61}{-1}{0}{5.6}
%         \myarrow{58}{63}{1}{-0.5}{5}
%     }
%     {{us_9978_standard_Free_100x100_sli0_R_10}
%     {37}}
%     {
%         \myarrow{59}{30}{-1}{-1.5}{4}
%         \myarrow{67}{33}{1}{-1.5}{4}
%     }
%     {54}

\appendix

\setcounter{figure}{0}
\renewcommand{\thefigure}{A\arabic{figure}}

\section{Patchification and De-patchification Operations}
\label{app:patch_operations}
During training, patches of size $64 \times 64 \times 8$ are randomly sampled from the dataset, with selection probabilities defined by a 2D truncated Gaussian distribution in space and a uniform distribution in time. This strategy preferentially samples patches from the central region of the image, which better captures the relevant anatomical features.

\begin{figure}[h]
    \centering

    % \includegraphics[width=0.9\linewidth]{figures/fig_patch_ramps.pdf}
    % \caption{Per-patch ramp weights for the running example
    % ($L = 64$, $O = 24$, $n = 3$). Each patch's weight profile depends on
    % whether it has a left and/or right neighbor.}
    % \label{fig:patch-ramps}

    % \vspace{0.8em}

    \includegraphics[width=0.9\linewidth]{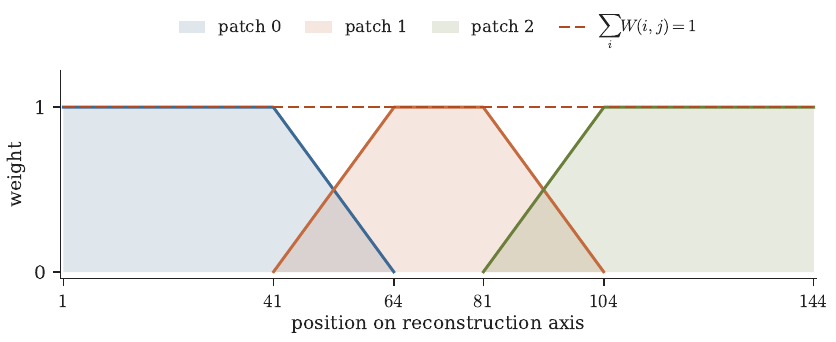}
    \caption{One-dimensional illustration of ramp-weighted averaging during de-patchification. Here, $i$ represents the patch index, $j$ represents the pixel location, and $W$ represents the relative weight of each patch at each overlapping pixel. In this example, the input length is $144$, the patch size is $64$, and the stride is $40$.}
    \label{fig:weights}
\end{figure}

During inference, the entire spatiotemporal image series is converted into overlapping patches of size $64 \times 64 \times 8$, with the number of patches kept as small as possible along each spatiotemporal dimension. As noted in Algorithm \ref{alg:patch_based_CineDiff}, inference requires both patchification $\mathcal{P}(\cdot)$ and de-patchification $\mathcal{P}^{-1}(\cdot)$ operations. For de-patchification, where the overlapping patches are converted back to the image series, we employ a ramp-weighted combination of the overlapped pixels. This scheme is illustrated in Figure \ref{fig:weights} for 1D. Compared to alternatives, including averaging across all neighboring patches in the overlapped region, the ramp-weighted average effectively eliminates boundary artifacts from the patches.

\setcounter{figure}{0}
\renewcommand{\thefigure}{S\arabic{figure}}
\section{Supplementary figures}
% ---------- Arrow settings for this figure only ----------
\def\ax{39}      % x position
\def\ay{56}      % y position
\def\adx{1}      % x direction
\def\ady{-1}      % y direction
\def\alen{4}    % arrow length

% ---------- Arrow 2 settings for this figure ----------
\def\bx{38}
\def\by{25}
\def\bdx{-1}
\def\bdy{-0.5}
\def\blen{5}

% ---------- Combine all arrows for this figure ----------
\def\figarrows{%
    \myarrow{\ax}{\ay}{\adx}{\ady}{\alen}%
    \myarrow{\bx}{\by}{\bdx}{\bdy}{\blen}%
}

% =========================
% Arrow settings for this figure only
% =========================
% ---------- Arrow 1' ----------
\def\axp{58}
\def\ayp{38}
\def\adxp{-1}
\def\adyp{0}
\def\alenp{10}

% ---------- Arrow 2' ----------
\def\bxp{61}
\def\byp{10}
\def\bdxp{-1}
\def\bdyp{0}
\def\blenp{10}

% ---------- Combine arrows ----------
\def\figarrowsyt{%
    \myarrow{\axp}{\ayp}{\adxp}{\adyp}{\alenp}%
    \myarrow{\bxp}{\byp}{\bdxp}{\bdyp}{\blenp}%
}

\begin{figure}[!htbp]%[t]
    \centering
    %------------ Row 1: images with titles on top ------------
    \begin{subfigure}[b]{0.23\textwidth}
        \centering
        \subcaption*{\large \sffamily Reference}
        \figuremetrics{PSNR}{SSIM}{LPIPS}{DISTS}
        \begin{overpic}[width=\linewidth]{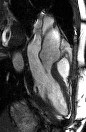}
            \figarrows
        \put(0,27){%
        \tikz{\draw[red!70,dashed,line width=1pt] (0,0) -- (3.79,0);}
        }
        \end{overpic}
    \end{subfigure}
    \hfill
    \begin{subfigure}[b]{0.23\textwidth}
        \centering
        \subcaption*{\large \sffamily CS}
        \figuremetrics{25.89}{0.802}{0.1774}{0.1522}
        \begin{overpic}[width=\linewidth]{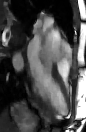}
            \figarrows
        \put(0,27){%
        \tikz{\draw[red!70,dashed,line width=1pt] (0,0) -- (3.79,0);}
        }
        \end{overpic}
    \end{subfigure}
    \hfill
    \begin{subfigure}[b]{0.23\textwidth}
        \centering
        \subcaption*{\large \sffamily CineVN}
        \figuremetrics{26.41}{0.844}{0.1326}{0.129}
        \begin{overpic}[width=\linewidth]{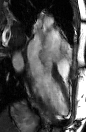}
        \put(0,27){%
        \tikz{\draw[red!70,dashed,line width=1pt] (0,0) -- (3.79,0);}
        }
            \figarrows
        \end{overpic}
    \end{subfigure}
    \hfill
    \begin{subfigure}[b]{0.23\textwidth}
        \centering
        \subcaption*{\large \sffamily CineDiff}
        \figuremetrics{27.80}{0.872}{0.1113}{0.1223}
        \begin{overpic}[width=\linewidth]{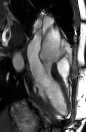}
        \put(0,27){%
        \tikz{\draw[red!70,dashed,line width=1pt] (0,0) -- (3.79,0);}
        }
            \figarrows
        \end{overpic}
    \end{subfigure}

    \vspace{0.5em}
    
    %------------ Row 2: error maps, no titles; ×10 on top-left ------------
    \begin{subfigure}[b]{0.23\textwidth}
        \centering
        \begin{overpic}[width=\linewidth]{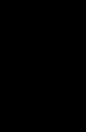}
        \end{overpic}
    \end{subfigure}
    \hfill
    \begin{subfigure}[b]{0.23\textwidth}
        \centering
        \begin{overpic}[width=\linewidth]{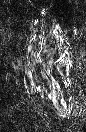}
            \put(0.5,93){\errlabel}
        \end{overpic}
    \end{subfigure}
    \hfill
    \begin{subfigure}[b]{0.23\textwidth}
        \centering
        \begin{overpic}[width=\linewidth]{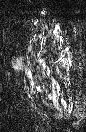}
            \put(0.5,93){\errlabel}
        \end{overpic}
    \end{subfigure}
    \hfill
    \begin{subfigure}[b]{0.23\textwidth}
        \centering
        \begin{overpic}[width=\linewidth]{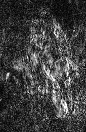}
            \put(0.5,93){\errlabel}
        \end{overpic}
    \end{subfigure}

    \vspace{0.5em}
    %------------ Row 3: time profile, no titles ------------
    \begin{subfigure}[b]{0.23\textwidth}
        \centering
        \begin{overpic}[width=\linewidth,height=2cm]{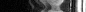}
        \figarrowsyt
        \end{overpic}
    \end{subfigure}
    \hfill
    \begin{subfigure}[b]{0.23\textwidth}
        \centering
        \begin{overpic}[width=\linewidth,height=2cm]{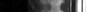}
        \figarrowsyt
        \end{overpic}
    \end{subfigure}
    \hfill
    \begin{subfigure}[b]{0.23\textwidth}
        \centering
        \begin{overpic}[width=\linewidth,height=2cm]{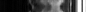}
        \figarrowsyt
        \end{overpic}
    \end{subfigure}
    \hfill
    \begin{subfigure}[b]{0.23\textwidth}
        \centering
        \begin{overpic}[width=\linewidth,height=2cm]{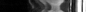}
        \figarrowsyt
        \end{overpic}
    \end{subfigure}
    
    \includegraphics[width=\textwidth]{figures/colorbar_gray.png}

    \caption{Reference image, reconstructions at $R=16$, and corresponding error maps (middle row) and time profiles (bottom row) along the dashed red line. The red arrows highlight details that are better preserved in CineDiff.}
    \label{fig:retro_P014_cine_lax_sli1_172x168_R_16}
\end{figure} % good, supplementary

\ReconTriplet
    {us_0011}{000}
    {CS, CineVN, and CineDiff reconstructions from a free-breathing RT prospectively undersampled dataset, along with the time profiles (bottom row) along the dashed red line. The red arrows highlight details that are better preserved in CineDiff.}
    % {Comparison of CS, CineVN, and CineDiff reconstructions for \texttt{us\_0011}, frame 000. RA: Move the arrows closer to the details.}
    {fig:us0011_f000}
    {trim=20 28 0 0} %L, D, R, U
    {
        \myarrow{44}{46}{1}{0}{6}
        \myarrow{28}{57}{1}{-1}{4}
    }
    {{us_0011_3T_sli0_100x144_sli0_R_9}
    {53}}
    {
        \myarrow{62}{8}{1}{1}{5.6}
        \myarrow{36}{34}{1}{-1}{5.6}
    }
    {49}

% \ReconTriplet
%     {us_9978}{000}
%     {CS, CineVN, and CineDiff reconstructions from a free-breathing RT prospectively collected from a human subject on a 0.55T scanner. The red arrows highlight details that are better preserved in CineDiff.}
%     % {Comparison of CS, CineVN, and CineDiff reconstructions for \texttt{us\_9978}, frame 000.}
%     {fig:us9978_f000}
%     {trim=15 35 10 10} %L, D, R, U
%     {
%         \myarrow{22}{61}{-1}{0}{5.6}
%         \myarrow{58}{63}{1}{-0.5}{5}
%     }
%     {{us_9978_standard_Free_100x100_sli0_R_10}
%     {37}}
%     {
%         \myarrow{59}{30}{-1}{-1.5}{4}
%         \myarrow{67}{33}{1}{-1.5}{4}
%     }
%     {54}

\ReconTriplet
    {us_9993}{000}
    {CS, CineVN, and CineDiff reconstructions from a free-breathing RT prospectively collected from a human subject on a 0.55T scanner, along with the time profiles (bottom row) along the dashed red line. The red arrows highlight details that are better preserved in CineDiff.}
    % {Comparison of CS, CineVN, and CineDiff reconstructions for \texttt{us\_9993}, frame 000.}
    {fig:us9993_f000}
    {trim=50 10 30 10} %L, D, R, U
    {
        \myarrow{49}{21}{-1}{-0.5}{5.6}
        \myarrow{38}{37}{-1}{-0.5}{5.6}
    }
    {{us_9993_standard_Free_100x100_sli0_R_10}
    {65}}
    {
        \myarrow{53}{58}{-1}{0}{7.6}
        \myarrow{62}{44}{-1}{0}{7.6}
    }
    {31}

\ReconTriplet
    {us_8997}{000}
    {CS, CineVN, and CineDiff reconstructions from a free-breathing RT prospectively collected from a porcine model on a 0.55T scanner, along with the time profiles (bottom row) along the dashed red line. The red arrows highlight details that are better preserved in CineDiff.}
        % {\\CS and MeanFlow reconstructions from a free-breathing RT\\ prospectively collected from a patient on a 3T scanner.\\ The red arrows highlight details that are better preserved in CineDiff.}
                % {\\Prospectively Undersampled\\Retrospectively Undersampled\\Reference, CS, and MeanFlow reconstructions from a breath-holding \\ retrospectively collected from a healthy volunteer on a 1.5T scanner.\\ The red arrows highlight details that are better preserved in CineDiff.}
    % {Comparison of CS, CineVN, and CineDiff reconstructions for \texttt{us\_8997}, frame 000.}
    {fig:us8997_f000}
    {trim=40 32 40 5} %L, D, R, U
    {
        \myarrow{31}{68}{1}{-1}{5.5}
        \myarrow{48}{30}{-1}{-1}{5.5}
    }
    {{us_8997_30kg_pig_sli2_Free_100x100_sli2_Free_R_10}
    {40}}
    {
        \myarrow{68}{48}{-1}{0}{8}
        \myarrow{60}{8}{1}{0}{8}
    }
    {42}

\end{document}